\documentclass[preprint2]{aastex63}

\usepackage{CJK}

\newcommand{\petit}{\texttt{petitRADTRANS}}

\newcommand{\pmn}{\texttt{PyMultiNest}}
\newcommand{\moog}{\texttt{MOOG}}

\received{\today}
\revised{}
\accepted{}
\submitjournal{ApJ}

\shorttitle{C/O HR 8799 c}
\shortauthors{Wang et al.}
\graphicspath{{./}{figures/}}

\begin{document}
\begin{CJK*}{UTF8}{gbsn}

\title{On the Chemical Abundance of HR 8799 and the Planet \lowercase{c}}

\correspondingauthor{Ji Wang}
\email{wang.12220@osu.edu}

\author[0000-0002-4361-8885]{Ji Wang (王吉)}
\affiliation{Department of Astronomy, The Ohio State University, 100 W 18th Ave, Columbus, OH 43210 USA}

\author[0000-0003-0774-6502]{Jason J. Wang (王劲飞)}
\altaffiliation{51 Pegasi b Fellow}
\affiliation{Department of Astronomy, California Institute of Technology, Pasadena, CA 91125, USA}

\author{Bo Ma}
\affiliation{School of Physics and Astronomy, Sun Yat-sen University, 2 Daxue Road, Zhuhai, Guangdong, 519082, People's Republic of China}

\author{Jeffrey Chilcote}
\affiliation{Department of Physics, University of Notre Dame, 225 Nieuwland Science Hall, Notre Dame, IN, 46556, USA}

\author{Steve Ertel}
\affiliation{Large Binocular Telescope Observatory, University of Arizona, 933 N Cherry Avenue, Tucson, AZ 85719, USA }
\affiliation{Steward Observatory, University of Arizona, 933 N Cherry Avenue, Tucson, AZ 85719, USA}

\author[0000-0002-1097-9908]{Olivier Guyon}
\affiliation{Subaru Telescope, NAOJ, 650 North A{'o}hoku Place, Hilo, HI 96720, USA}
\affiliation{Steward Observatory, University of Arizona, 933 N Cherry Avenue, Tucson, AZ 85719, USA}
\affiliation{Astrobiology Center of NINS, 2-21-1 Osawa, Mitaka, Tokyo 181-8588, Japan}

\author{Ilya Ilyin}
\affiliation{Leibniz-Institute for Astrophysics Potsdam (AIP), An der Sternwarte 16, D-14482 Potsdam, Germany}

\author[0000-0001-5213-6207]{Nemanja Jovanovic}
\affiliation{Department of Astronomy, California Institute of Technology, MC 249-17, 1200 E. California Blv, Pasadena, CA 91106 USA}

\author{Paul Kalas}
\affiliation{Department of Astronomy, University of California, Berkeley, CA 94720, USA}
\affiliation{SETI Institute, Carl Sagan Center, 189 Bernardo Ave.,  Mountain View CA 94043, USA}

\author{Julien Lozi}
\affiliation{Subaru Telescope, NAOJ, 650 North A{'o}hoku Place, Hilo, HI 96720, USA}

\author{Bruce Macintosh}
\affiliation{The Kavli Institute for Particle Astrophysics and Cosmology, Stanford University, Stanford, CA 94305, USA}

\author{Jordan Stone}
\altaffiliation{Hubble Fellow}
\affiliation{Steward Observatory, University of Arizona, 933 N Cherry Avenue, Tucson, AZ 85719, USA}

\author[0000-0002-6192-6494]{Klaus G. Strassmeier}
\affiliation{Leibniz-Institute for Astrophysics Potsdam (AIP), An der Sternwarte 16, D-14482 Potsdam, Germany}
\affiliation{Institute for Physics and Astronomy, University of Potsdam, Karl-Liebknecht-Str. 24/25, D-14476 Potsdam, Germany}



\begin{abstract}

Comparing chemical abundances of a planet and the host star reveals the origin and formation path. Stellar abundance is measured with high-resolution spectroscopy. Planet abundance, on the other hand, is usually inferred from low-resolution data. For directly imaged exoplanets, the data are available from a slew of high-contrast imaging/spectroscopy instruments. Here, we study the chemical abundance of HR 8799 and its planet c. We measure stellar abundance using LBT/PEPSI (R=120,000) and archival HARPS data: stellar [C/H], [O/H], and C/O are 0.11$\pm$0.12, 0.12$\pm$0.14, and 0.54$^{+0.12}_{-0.09}$, all consistent with solar values. 

We conduct atmospheric retrieval using newly obtained Subaru/CHARIS data together with archival Gemini/GPI and Keck/OSIRIS data. We model the planet spectrum with \petit\ and conduct retrieval using \pmn. Retrieved planetary abundance can vary by $\sim$0.5 dex, from sub-stellar to stellar C and O abundances. The variation depends on whether strong priors are chosen to ensure a reasonable planet mass. Moreover, comparison with previous works also reveals inconsistency in abundance measurements. We discuss potential issues that can cause the inconsistency, e.g., systematics in individual data sets and different assumptions in the physics and chemistry in retrieval. We conclude that no robust retrieval can be obtained unless the issues are fully resolved.   

\end{abstract}



\section{Introduction}
\label{sec:intro}

{{HR 8799 bcde remain one of only a few multi-planet systems that have been directly imaged~\citep{Marois2008, Marois2010}, the other two systems being PDS 70 b and c~\citep{Keppler2018, Haffert2019} and $\beta$ Pic b~\citep{Lagrange2010} and c~\citep{Lagrange2019}}}. The HR 8799 system has been extensively studied previously. The astrometry of the four planets are measured to a few milliarcsecond precision~\citep[e.g.,][]{Konopacky2016, Wertz2017}. The atmospheres of planets in this system have been studied by multi-band photometry and low-resolution spectroscopy~\citep[e.g.,][]{Skemer2014, Bonnefoy2016, Zurlo2016, Lavie2017}.

Planet c is the most amenable for current-generation high contrast instruments for its favorable separation and planet-star flux ratio. Planet d and e are so close to the host star so that speckle noise significantly compromises data quality. The planet-star flux ratio of planet b is lower than that of c despite a larger separation. 

Consequently, rich data sets have been obtained for HR 8799 c, including integral field spectroscopy (IFS) data from Palomar/P1640~\citep{Oppenheimer2013}, Keck/OSIRIS~\citep{Konopacky2013}, and Gemini-S/GPI~\citep{Ingraham2014, Greenbaum2018}. {{Readers are referred to~\citet{Bonnefoy2016, Zurlo2016} and subsequent references~\citep{Greenbaum2018,Wang2018,Ruffio2019,Gravity2019,Petit2020} for a summary of observations of HR 8799 planets.}} VLT/SPHERE has a too small a field of view (FOV) for IFS data for HR 8799 c. At high spectral resolution,~\citet{Wang2018} obtained $L$-band Keck/NIRSPEC (R=15,000) data and detected water in the atmosphere of HR 8799 c. In the near-term future, it is expected that high quality data will come from JWST and new-generation high-contrast imaging and spectroscopy instruments~\citep[e.g., ][]{Mawet2017, Currie2018,Skemer2018,Gravity2019}. 

Spectral analysis and atmospheric retrieval have been applied to HR 8799 c. Multiple lines of evidence suggest that the atmosphere is cloudy and not in chemical equilibrium. Evidence from photometric data includes the suppressed $J$ band flux and CH$_4$ narrow-band absorption in $L$ band~\citep{Skemer2014}. Spectral fitting and atmospheric retrieval also suggest a cloudy atmosphere and significant vertical mixing that causes chemical disequilibrium~\citep{Madhusudhan2011, Konopacky2016, Lavie2017}. The C/O ratio has been constrained to inform the formation and accretion history of the planet~\citep{Konopacky2013, Lavie2017}. 

Despite much progress on abundance measurements, there are two outstanding questions. First, how well can we constrain the stellar abundance? It is the comparison between stellar and planetary abundances that reveals the pathway of planet formation. Previous works used stellar abundances from~\citet{Sadakane2006}, which is based on data of spectral resolution of 42,000. Can the abundance measurements be improved and the associated uncertainties be properly accounted for? 

Second, how can we access the robustness of atmospheric retrieval and its implications to planet formation? More specifically, how do we develop a retrieval framework that combines multiple data sets, how do we interpret the discrepancy between different data sets, and how do we properly set priors to ensure a physically and chemically sensible retrieval?

To better measure stellar abundances, we obtained high signal-to-noise ratio (SNR$\sim$400) and high-spectral-resolution (R=120,000) data using the Potsdam Echelle Polarimetric and Spectroscopic Instrument (PEPSI) at the large binocular telescope (LBT)~\citep{Strassmeier2015}, and used archival HARPS (R=115,000) data~\citep{Mayor2003}. 

To improve planet atmospheric retrieval, we supplemented new IFS data to existing Gemini/GPI and Keck/OSIRIS data. Our new data is from the Coronagraphic High Angular Resolution Imaging Spectrograph (CHARIS) that simultaneously covers $J$, $H$, and $K$ band~\citep{Peters2012}. We used newly-developed atmospheric modeling package \petit~\citep{Molliere2019} and \pmn~\citep{Buchner2014} to sample posterior distributions. 




\begin{figure*}[ht]
\epsscale{1.0}
\plotone{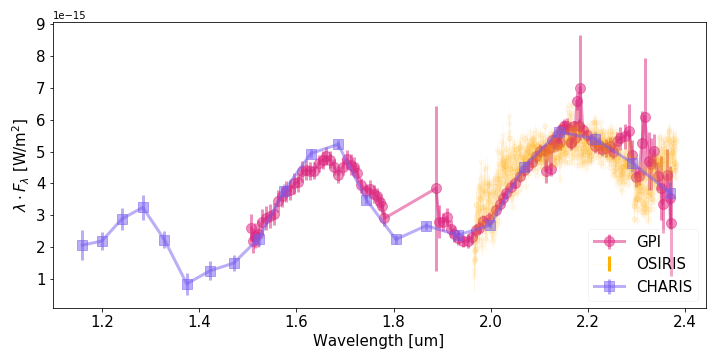}
\caption{Data sets used in this work are from CHARIS, GPI, and OSIRIS.  
\label{fig:dataset}}
\end{figure*} 

The paper is organized as follows. \S \ref{sec:observation} summarizes the observational data. \S \ref{sec:st_ab} describes stellar abundance measurements and results. \S \ref{sec:data_reduction} describes the framework of planet atmosphere retrieval and results are given in \S \ref{sec:combined_retrieval}. Discussion and summary are given in \S \ref{sec:dis} and \S \ref{sec:summary}.

\section{Data}
\label{sec:observation}

We used high-resolution spectra to determine stellar abundances. For planet atmospheric abundances, we used IFS data sets for HR 8799 c. While photometric data points are valuable and can be easily taken into account in our analysis, it is unclear how systematics in photometry would affect our analysis, e.g., whether the systematics is astrophysical, coming from the instrument, or from data reduction. 

We supplemented new observational data to archival data. Below we provide details on the data that we used for stellar and planetary abundance measurements in two categories: new and archival data. 

\subsection{New Data}

\subsubsection{LBT/PEPSI Observation of HR 8799}

Large Binocular Telescope (LBT, $2\times 8.4$\,m on Mt.\ Graham, Arizona, USA) observations with the PEPSI spectrograph~\citep{Strassmeier2015} were made on November 14, 2019 UT with a 200\,$\mu$m fiber ($R=120\,000$) in two spectral regions 4265 - 4800\,\AA\ and 5441 - 6278\,\AA\ (cross-dispersers 2 and 4) with 5\,min integration time starting at 02:16:52 UT. {{The telescope time belongs to an LBTI~\citep{Hinz2016} engineering night (PI: Stone)}}, the telescope was pointed to the star with binocular mode with PEPSI on one side (right DX) and LBTI on the other one ({{Left SX}}).  The data were processed as described in~\citet{Strassmeier2018} and yielded the maximal signal-to-noise ratio of 340 and 500 in two cross-dispersers.

\subsubsection{Subaru/CHARIS Observation of HR 8799 c}
HR 8799 was observed on two consecutive nights, 2018 September 1 and 2018 September 2 (PIs: Jason Wang and James Graham), using the CHARIS integral field spectrogrpah \citep{Groff2015, Groff2017} behind the SCExAO adaptive optics system \citep{Jovanovic2015}. A full analysis of the data will be published in a following paper (Wang et al. in prep), but here we will briefly summarize the observations and reduction relevant for spectroscopy of HR 8799 c. With the broadband mode of CHARIS, we obtained R$\sim$20 integral field spectrscopy of HR 8799 c from 1.1 to 2.4 $\mu$m, covering $J$, $H$, and $K$ bands simultaneously in 22 spectral channels. In this work, we used data corresponding to 6.45~hours and 6.72~hours of open shutter time data on HR 8799 c from night 1 and night 2 respectively. 

The raw integral field spectroscopy data were turned into spectral data cubes using the CHARIS data reduction pipeline \citep{Brandt2017}. Stellar PSF subtraction and spectral extraction was performed using the open-source package \texttt{pyKLIP} \citep{Wang2015_pyklip} that implements the forward modeling framework presented in \citep{Pueyo2016}. Uncertainties and biases from the spectral extraction were estimated using simulated planets injected into the data at the same separation as HR 8799 c, but at different azimuthal positions. The data are flux calibrated using artificial satellite spots injected into each exposure \citep{Jovanovic2015_satspots}. From binary calibrators {{taken during the night}}, we determined the satellite spots to have a flux ratio relative to the star of $2.64 \times 10^{-3} \times (\lambda/1.55~\mu m)^{-2}$, where $\lambda$ is the wavelength of each spectral channel {{(Wang et al. in prep)}}.

CHARIS data of two nights were combined in the following way. We first flux-calibrated the data for each night using GPI $H$ and $K$-band IFS data. Then the average flux of the two nights was taken as the flux. Error bars are the larger of the following: (1) summed quadrature of error bars that were reported by CHARIS data reduction pipeline, and (2) half of the difference of flux measurements from the two nights.  

\subsection{Archival Data}



\subsubsection{ESO-3.6m/HARPS data for HR 8799}
Data of HR~8799 using the HARPS spectrograph from the 3.6 m ESO telescope were taken in 2009 (PI: Chauvin, ESO runs Program ID: 083.C-0794), with a wavelength coverage of 3800$-$6900~\AA\ and spectral resolution of $\sim$115,000. The data were reduced by the standard HARPS pipeline. Eight 180s exposures were reduced and combined to yield a one dimensional spectrum for spectral analysis, which has a SNR (per pixel) of $\sim$300 at 5500~\AA. 

\subsubsection{Keck/OSIRIS data for HR 8799 c}
OSIRIS $K$-band IFS data (R$\sim$5000) were reported in~\citet{Konopacky2013}, in which absorption lines of molecular species such as H$_2$O and CO were detected using a cross-correlation technique. The higher spectral resolution allowed for an investigation of atmospheric C/O and probed the formation and accretion history of HR 8799 c~\citep{Oberg2011, Madhusudhan2012}. 

\subsection{Gemini-S/GPI data for HR 8799 c}
Newly analyzed GPI data from 1.5 $\mu$m to 2.4 $\mu$m (R$\sim$45-80) were reported in~\citet{Greenbaum2018}, where they found significant discrepancy between the newly-analyzed and previously-analyzed data~\citep{Ingraham2014}. The discrepancy was attributed to over subtraction in spectral extraction. 

\section{Stellar Atmospheric Parameters and Abundances of HR 8799}
\label{sec:st_ab}
\subsection{Stellar Parameters}
We derived the stellar atmospheric parameters of HR~8799 from the PEPSI and HARPS spectra using (1) the spectral analysis package \moog\ \citep{Sneden1973}, (2) a line list consisting of 48 Fe I and 18 Fe II lines, and (3) ATLAS9 stellar atmospheric grid models. The equivalent widths (EWs) of the spectral lines were measured using TAME \citep{kang12}, which used a Gaussian profile to fit the absorption lines. This EW technique is based on the iron ionisation equilibrium and excitation equilibrium. We require that (1) Fe I lines and Fe II lines give the same averaged [Fe/H] abundance, (2) no correlations between [Fe I/H] values and reduced EW values, and (3) no correlation between [Fe I/H] and excitation potential of the lines. The errors are estimated using the same method as that in \citet{Tabernero2019}. We used [ ] to denote logarithmic abundance with respect to solar value~\citep{Asplund2009}.

The initial guesses of the stellar parameters were taken from \citet{Sadakane2006} and then iterated on by slightly changing each parameter until above requirements were met. The final adopted stellar atmospheric parameters are $T_{\rm eff}=7390\pm80$ K, $\log (g)=4.35\pm0.07$, $\rm [Fe/H]=-0.52\pm0.08$ and $\rm \epsilon=3.1\pm0.2~km~s^{-1}$, as shown in Table.~\ref{tab:stellarparams}.

\subsection{C and O Abundances}
We then proceeded to derive the abundances of C and O using the HARPS data. Adopting the stellar atmosphere models of Kurucz ATLAS9 with the above derived atmospheric parameters, the abundances of C and O were derived using the \texttt{abfind} driver from \moog, results of which are shown in Table.~\ref{tab:CO_abundances}. The errors associated with abundance measurements were calculated by accounting for the atmospheric parameter errors presented in Table.~\ref{tab:stellarparams} and the measurement errors of the EWs. The abundance of C was derived from C I 5052.2, 4771.7, 4932.0, and 5380.3~\AA\ features. The abundance of O was derived from the O I triplet feature at 6156$–$6158~\AA, with their line values taken from the NIST database. 

We adopt the values from~\citet{Asplund2009}. For example, [C/H] = 0 corresponds to $\log$(C/H) = -3.57, [O/H] = 0 corresponds to $\log$(O/H) = -3.31, and [C/O] = 0 corresponds to C/O = 0.55. Uncertainty of solar C and O values is 0.05 dex. However, we note that solar values reported in~\citet{Palme2014} are somewhat different: $\log$(C/H) = -3.50$\pm$0.06, $\log$(O/H) = -3.27$\pm$0.07, and C/O = 0.59.  

The final adopted [C/H] and [O/H] were $0.11\pm0.12$~dex and $0.12\pm0.14$~dex respectively, where the solar abundance values were taken from \citet{Asplund2009}. The ratio of C/O was 0.54$^{+0.12}_{-0.09}$, where the error from uncertainty of stellar parameters was small because both C and O abundances change in the same direction as the effective temperature of the stellar model changes. 


The abundances were estimated based on a LTE model. The non-LTE correction for C and O abundances are small ($\sim0.04$~dex) for HR~8799 based on the estimate of \citet{Takeda1999} for late-A type stars. 
\subsection{Comparing to Previous Works}

{{HR~8799 has been identified as a $\lambda$~Bootis type star~\citep{Gray1999}. A $\lambda$~Bootis type star has solar surface abundances for volatile elements such as C, N, O and S, and sub-solar surface abundances of Fe-peak elements. The potential origins of this type of stars are discussed in \S \ref{sec:lambdaBootis}, and one explanation could be related to the debris disk~\citep{Draper2016} and planets~\citep{Jura2015} around HR 8799. HR 8799 is given the following stellar parameters: $T_{\rm eff}=7424$ K, $\log (g)=4.22$, and $\rm [Fe/H]=-0.5$, using a spectrum with a low spectral resolution~\citep[$\Delta\lambda=1.8~\AA$,][]{Gray2003}. }}

By using high-resolution spectra from the Elodie spectrograph (R=42,000), \citet{Sadakane2006} derived the stellar atmospheric parameters and chemical abundances of HR~8799. The results are consistent within uncertainties although we used a different set of line lists when deriving the atmospheric parameters. As for the C and O abundances, we found relatively lower values than those reported in \citet{Sadakane2006}. Their values are $\log{\epsilon_{\rm C}}=8.63$ and $\log{\epsilon_{\rm O}}=8.88$, in comparison to our values in Table \ref{tab:CO_abundances}: $\log{\epsilon_{\rm C}}=8.54$ and $\log{\epsilon_{\rm O}}=8.81$. The reason behind this is likely due to a different set of stellar atmospheric model are used. The C/O ratio is 0.54$^{+0.12}_{-0.09}$ from this work, comparing with 0.56 from \citet{Sadakane2006}.

\section{Atmosphere Modeling and Retrieval for HR 8799 \lowercase{c}}
\label{sec:data_reduction}

\subsection{Modeling Planet Atmospheres}
We used \petit~to model exoplanet atmospheres~\citep{Molliere2019}. \petit~is a versatile package that can model transmission and thermal spectra at high {{(R=1,000,000)}} and low spectral (R=1000) resolution with opacities available for all major chemical species and their isotopic species. We modeled the thermal spectrum at low resolution, considering four molecular species including H$_2$O, CO, CO$_2$, and CH$_4$. In addition, Rayleigh scattering due to H$_2$ and He, and collisional broadening due to H$_2$-H$_2$ and H$_2$-He, were considered. 

\subsection{Parameterization}
\label{sec:parametrization}
The following input parameters were used for \petit: planet radius, mass mixing ratio for H$_2$O, CO, CO$_2$, and CH$_4$. We included the corresponding pressure at which an optically-thick global cloud forms. In addition, we used analytical pressure-temperature (P-T) profile~\citep{Parmentier2014, Parmentier2015}. {{For directly-imaged exoplanets with low stellar irradiation levels, the P-T profile is nearly iso-thermal at high altitudes where energy transport is radiative and adiabatic at depth (e.g., optical depth $>$ 1} for convective atmospheres~\citep{Parmentier2015}. Parameters to define a P-T profile were: surface gravity and internal temperature (t$_{\rm{int}}$), which describes the heat flux from the planet interior}. Potential correlated noise in these data sets was modeled using Gaussian process with three parameters: correlation length, correlation amplitude, and white noise amplitude.

To facilitate comparison with observations, we also calculate planet luminosity and effective temperature. Planet luminosity is calculated by integrating retrieved planet spectral energy distribution from 0.5 to 15 $\mu$m, where the bulk of planet emission is. Planet effective temperature ($t_{\rm{eff}}$) is then inferred using the following equation: $L=4\pi{\rm{R}}_P^2\cdot\sigma t_{\rm{eff}}^4$, where ${\rm{R}}_P$ is planet radius and $\sigma$ is the Stephan-Boltzmann constant. 

\subsection{Chemistry}
We calculated C/H with the following equation:
\begin{equation}
\label{eq:extraction}
C/H = \frac{X_{\rm{CO}} + X_{\rm{CO_2}} + X_{\rm{CH_4}}}{2 \times X_{\rm{H_2}} + 4 \times X_{\rm{CH_4}} + 2 \times X_{\rm{H_2O}}},
\end{equation}
where X is volume mixing ratio. The conversion from mass mixing ratio, which was used in model parameters, to volume mixing ratio is as follows: 
\begin{equation}
\label{eq:extraction}
X_{i} = {\rm{{mr}}}_i \times {\rm{MMW}} / \mu_{i},
\end{equation}
where ${{\rm{mr}}}_{i}$ is mass mixing ratio and subscript $i$ denote a molecular species, $\mu$ is molecular weight in atomic mass unit, and MMW is mean molecular weight, which is defined as:
{
\begin{equation}
\label{eq:extraction}
{\rm{MMW}} = \left(\sum_{i} \frac{\rm{mr}_{i}}{\mu_{i}}\right)^{-1}.
\end{equation}
}
The mass mixing ratios for all considered species add up to unity. Molecular hydrogen and helium ratio is 3:1, which is based on the assumption of primordial composition. Similarly, O/H was calculated using the following equation:
\begin{equation}
\label{eq:extraction}
O/H = \frac{X_{\rm{CO}} + 2 \times X_{\rm{CO_2}} + X_{\rm{H_2O}}}{2 \times X_{\rm{H_2}} + 4 \times X_{\rm{CH_4}} + 2 \times X_{\rm{H_2O}}}.
\end{equation}
And C/O was calculated as:
\begin{equation}
\label{eq:extraction}
C/O = \frac{X_{\rm{CO}} + X_{\rm{CO_2}} + X_{\rm{CH_4}}}{X_{\rm{CO}} + 2 \times X_{\rm{CO_2}} + X_{\rm{H_2O}}}.
\end{equation}

\subsection{Retrieval}
We used PyMultiNest~\citep{Buchner2014} for posterior sampling. Priors are listed in Table \ref{tab:prior} and the likelihood function is $\exp[{-(\mathcal{D}-\mathcal{M})^2/\mathcal{E}^2}]$, where $\mathcal{D}$ is data, $\mathcal{M}$ is model, and $\mathcal{E}$ is the error term.

\section{Retrieval Results}
\label{sec:combined_retrieval}

We report our atmospheric retrieval results based on a combined data set that includes GPI, CHARIS, and OSIRIS data. As shown in the Appendix, retrieval based on individual GPI and CHARIS data returns reasonable agreement between modeled spectra and data points, so we use the GPI and CHARIS data as they are. In comparison, OSIRIS data have systematics that cannot be accounted for by our model. We first provide detailed treatment of the OSIRIS data before working on the combined data set. Subsequently, we will show that the choice of priors significantly affects the retrieval results, leaving a cautionary note on atmospheric retrieval.

\subsection{Using Normalized OSIRIS Data}

OSIRIS data have a higher spectral resolution (R=5,000) than the low-res mode of \petit\ (R=1,000). We could in principle use the high-res mode of \petit\ (R=1,000,000), but the computational time for the whole $K$ band wavelength coverage prevents us from sampling the parameter space in a reasonable amount of time. We therefore decide to down-sample the OSIRIS data to R=1,000 to match the low-res mode of \petit. While this procedure potentially degrades the original data, spectral features such as lines and band heads would be mostly preserved even at R=1,000.

As shown in \S \ref{sec:osiris_result}, we cannot obtain a reasonable agreement between our modeled spectra and the absolute flux measured from OSIRIS. We suspect that OSIRIS data have some systematic error that alters its spectral shape, which results in the challenges in using OSIRIS absolute flux. {{Our suspicion is further supported by the agreement between GPI data and the fitted modeled spectra (\ref{sec:gpi_result}), which suggests that the modeled spectra can reasonably fit certain data sets with less contribution from systematic errors. }} If systematics only affects low-frequency part of the spectrum, e.g., overall spectral shape etc., then we can remove the low-frequency systematics with a high-pass filter and use the normalized flux for retrieval. The normalized flux was also used in~\citet{Konopacky2013}. 

\subsection{Weak Priors}
\label{sec:weak_priors}
\begin{figure}[h!]
\epsscale{1.0}
\plotone{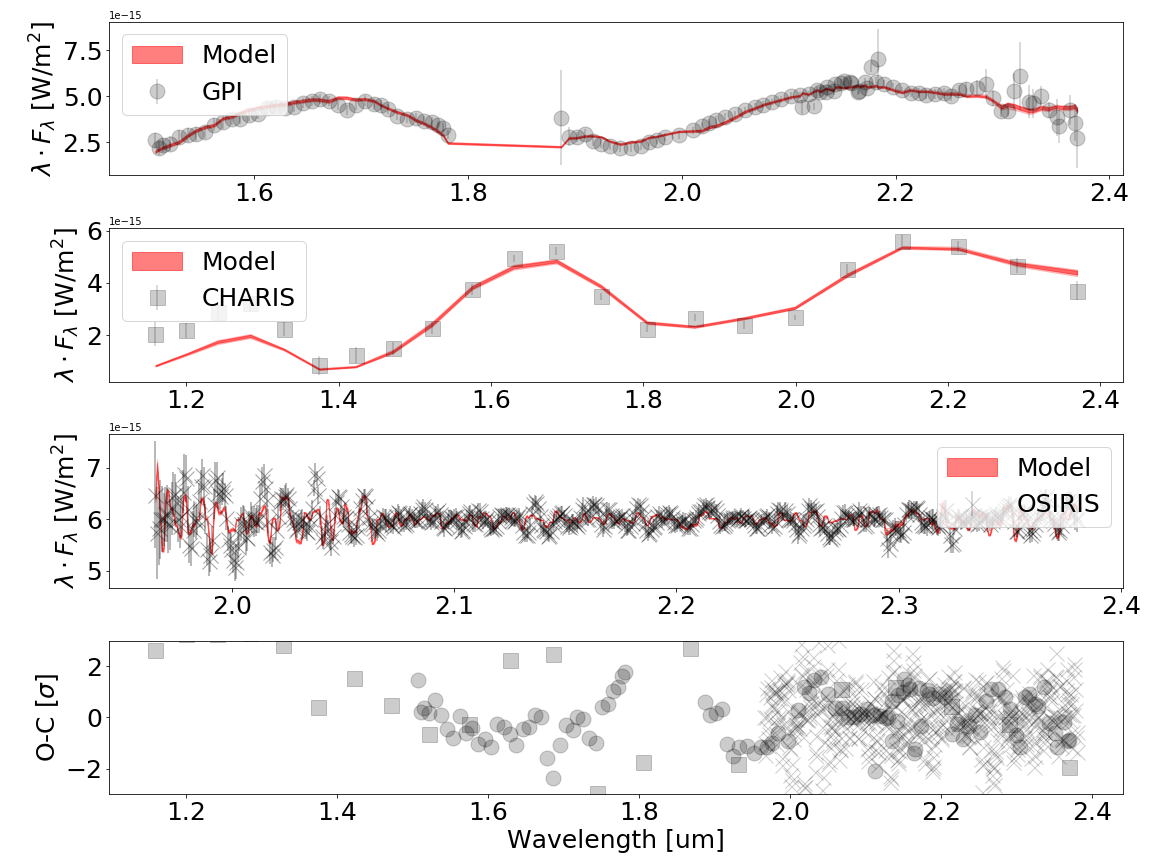}
\caption{Top three panels are data and modeled spectra for GPI (filled circles), CHARIS (squares), and OSIRIS (crosses, normalized). Bottom: residual plot with data minus model and divided by errors. 
\label{fig:gco_w_data_model}}
\end{figure} 

Spectra based on retrieval posterior samples are shown in Fig. \ref{fig:gco_w_data_model}. Modeled spectra trace data points reasonably well except for $J$-band: models underpredict planet flux when compared to CHARIS data points. 

As shown in Table \ref{tab:mcmc_result}, planet luminosity and effective temperature are within the range from~\citet{Konopacky2013}, i.e., 900-1300 K and $-4.8<\log(L/L_\odot)<-4.6$. We note here the difference between t$_{\rm{int}}$ and t$_{\rm{eff}}$. T$_{\rm{eff}}$ is generally used in grid model of spectra. T$_{\rm{int}}$ is parameter used in P-T profile {{that describes the heat flux from the planet interior}} (see \S \ref{sec:parametrization}). A higher t$_{\rm{int}}$ does not necessarily translate into a higher t$_{\rm{eff}}$. This is because there are other factors that regulate the emerging flux, e.g., cloud and molecular opacities. 

The retrieved gravity $\log g$ is 3.97$\pm$0.03 and radius is 1.47$\pm$0.02 R$_{\rm{Jupiter}}$. The planet mass would be 8.1$\pm$0.6 M$_{\rm{Jupiter}}$. While it is consistent with the estimate from~\citep{Wang2018b}, it is slightly higher than the previously allowed maximum 7 M$_{\rm{Jupiter}}$ from a dynamical stability point of view~\citep{Marois2010, Fabrycky2010}. 

\subsection{Strong Priors}
\label{sec:strong_priors}

\begin{figure}[h!]
\epsscale{1.0}
\plotone{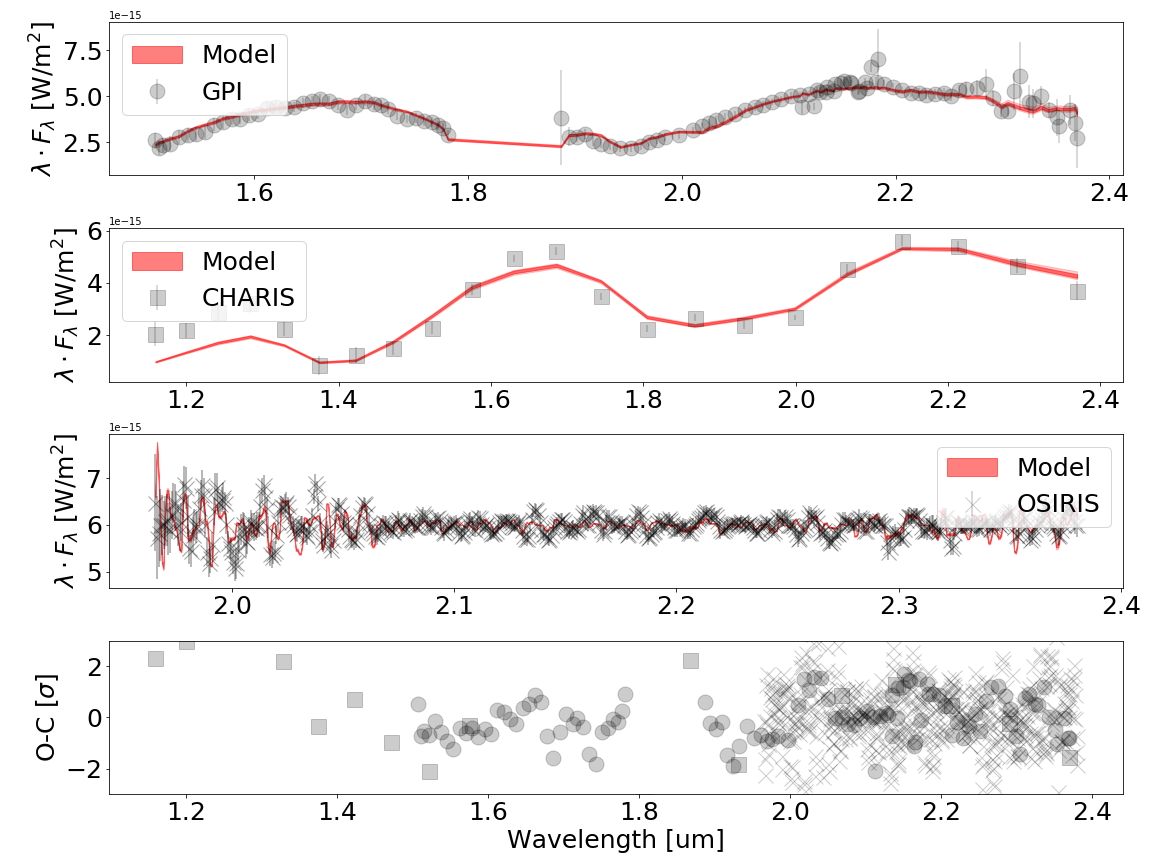}
\caption{Same as Fig. \ref{fig:gco_w_data_model} but with strong priors on planet temperature and radius.  
\label{fig:gco_s_data_model}}
\end{figure} 

Given the aforementioned issue in inferred planet mass, we decide to put stronger priors to restrain the possible parameter space for posterior samples. We set the priors for internal temperature and planet radius to be Gaussian functions with negligible standard deviation. This effectively set the internal temperature and planet radius to be 1200 K and 1.2 R$_{\rm{Jupiter}}$. The t$_{\rm{int}}$ of 1200 K is chosen to be more consistent with retrieved t$_{\rm{int}}$ using GPI or CHARIS individual data set, which shows better agreement between models and data (see Table \S \ref{tab:mcmc_result} and \S \ref{app:individual_analysis}). 

At the retrieved $\log g=3.96$, the corresponding planet mass is 5.28 M$_{\rm{Jupiter}}$. This is a reasonable mass given the dynamical stability constraint~\citep{Fabrycky2010}. The comparison between the weak and strong prior cases highlights the limitation of retrieval: the most likely parameter space to fit the data may not be physically/chemically allowed. 

Because of the different priors, the retrieved posterior distributions are significantly different for different parameters. Specifically, in order to maintain the same flux, the weak-prior case returns higher abundances and thus high opacity to offset the higher internal temperature (see Table \ref{tab:mcmc_result}). 

\subsection{Adding $L$- and $M$-band Photometric Data}
\label{sec:lm}

\begin{figure}[h!]
\epsscale{1.0}
\plotone{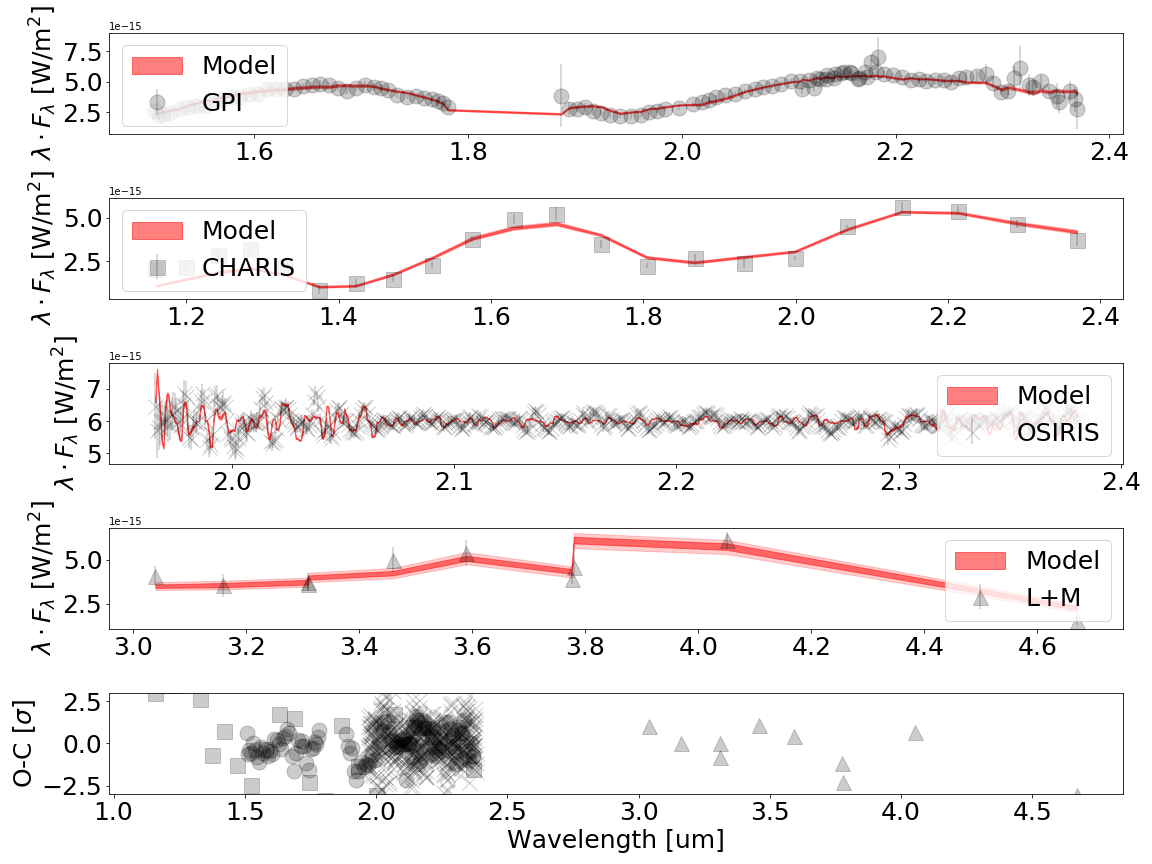}
\caption{Top four panels are data and modeled spectra for GPI (filled circles), CHARIS (squares), OSIRIS (crosses, normalized), and $L$- and $M$-band photometric data (triangles). Bottom: residual plot with data minus model and divided by errors.  
\label{fig:gco_lm_data_model}}
\end{figure} 

Given the significantly different results between the weak-prior and the strong-prior cases, we explore if adding additional constraints would reconcile the difference. We include $L$- and $M$-band photometric data that are complied by~\citet{Bonnefoy2016} from previous works~\citep{Galicher2011, Skemer2012, Currie2014, Skemer2014}. With weak priors, the retrieval results are shown in Fig. \ref{fig:gco_lm_data_model} and summarized in Table. \ref{tab:mcmc_result}. 

Dynamical stability is no longer a concern because the retrieved planet radius (1.05 R$_{\rm{Jupiter}}$) and surface gravity ($\log g=3.99$) are consistent with a 4.4 jupiter-mass planet, an even lighter planet than the strong-prior case. As shown in the section below, adding $L$- and $M$-band photometric data represents a solution that is in between the weak-prior and the strong-prior cases.   

\subsection{C/O}
\label{sec:co}
\begin{figure*}[h!]
\epsscale{1.0}
\plotone{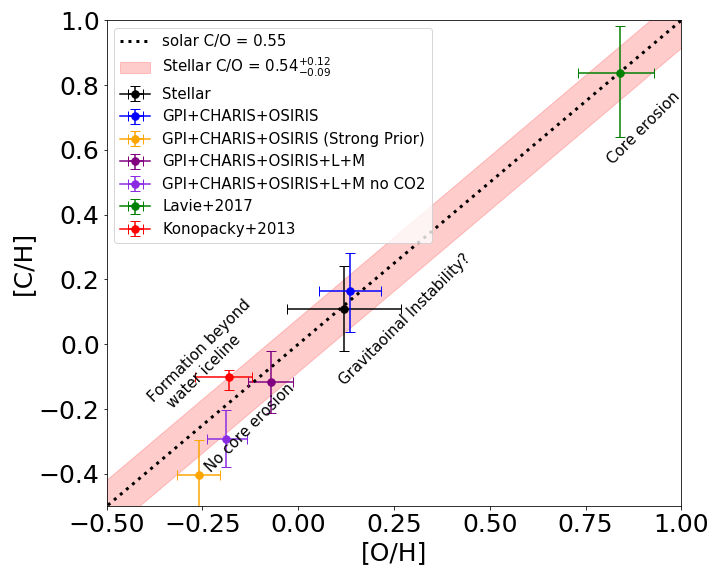}
\caption{Carbon and oxygen abundance for HR 8799 (black) and planet c (colored data points). Solid and dashed lines represent solar C/O~\citep[0.55, ][]{Asplund2009} and the C/O for HR 8799 (0.38). Red shaded region is the 1-$\sigma$ uncertainty region for HR 8799 C/O. 
\label{fig:co}}
\end{figure*} 

C and O are two elements that can be measured in exoplanetary atmospheres and shed light on planet formation. Fig. \ref{fig:co} shows our measurement of C and O abundance on the [C/H] - [O/H] plane, where brackets mean logarithmic abundance ratio with respect to the solar value. Such a [C/H] - [O/H] plot can be used to diagnose planet formation history~\citep{Madhusudhan2012, Madhusudhan2017}. 

{{Our results, depending on choice of priors, imply different formation path for HR 8799 c (Table \ref{tab:CO_abundances}). The result with weak priors shows consistent C and O abundances with the star. The interpretation could be (1), formation through direct gravitational instability; (2), accretion of carbon-enriched planetesimals (e.g., tar) to explain the slightly higher than stellar C abundance; and/or (3), core erosion after planet formation. The result with strong priors is concerning because there is no plausible formation scenario in which a planet has both sub-stellar C and O abundance and sub-stellar C/O according to \citet{Madhusudhan2017}. }}

The data point corresponding to the case after including $L$- and $M$-band photometric data is located in between the weak- and strong-prior cases. In addition, it is consistent with the data point based on the result from~\citet{Konopacky2013}. The result shows a planet C/O that consistent with stellar C/O. Both C/H and O/H are sub-stellar. This data point is in line with a scenario of formation beyond water iceline and no core erosion. {{The water iceline is located at $\sim$3-10 AU from the host star assuming an optical thin disk~\citep{Lavie2017}.}}

\subsection{Comparing results with Previous Works}
More intriguingly, results from previous works~\citep{Konopacky2013, Lavie2017} also show a significant difference. In summary, the four data points are roughly consistent with stellar C/O, but vary greatly in [C/H] and [O/H]. 

The data point for~\citet{Lavie2017} shows super-stellar C and O abundances and stellar C/O, implying that (1) the planet core form beyond the CH$_4$ ice line where C/O is at the stellar value; and (2) over time, core erosion place a role in changing C/H and O/H from sub-solar to super-solar. We note that the data point is at the maximum possible C/H and O/H enrichment from core accretion, i.e., a factor of $\sim$4 based on~\citet{Madhusudhan2017}.

The data point for~\citet{Konopacky2013} shows a slightly super-stellar C/O and sub-stellar C/H and O/H values. Since there is not posterior distribution, we estimate the error bar for this data point based on the quoted C/O value (0.65$^{+0.10}_{-0.05}$) and the corresponding C/H and O/H values provided in Table S1 in~\citet{Konopacky2013}. The most likely explanation, based on~\citet{Madhusudhan2017}, is that the planet form between the H$_2$O and CO$_2$ ice line, accreting gas that is deficient of H$_2$O, and core erosion plays no role in enriching the planet atmosphere. 

\section{Discussion}
\label{sec:dis}
\subsection{Comparing to~\citet{Lavie2017}}

\citet{Lavie2017} conducted comprehensive study on atmospheric retrieval for the four planets in HR 8799 system. Their methodology shares some similarities to ours, e.g., using multiple data sets and using \pmn\ for posterior sampling, although there are several key differences that will be discussed below. A comparison of our method and their approach would shed light on how to reconcile the difference and improve retrieval for future data sets.    

\subsubsection{Common Parameters}
\begin{figure}[h!]
\epsscale{1.0}
\plotone{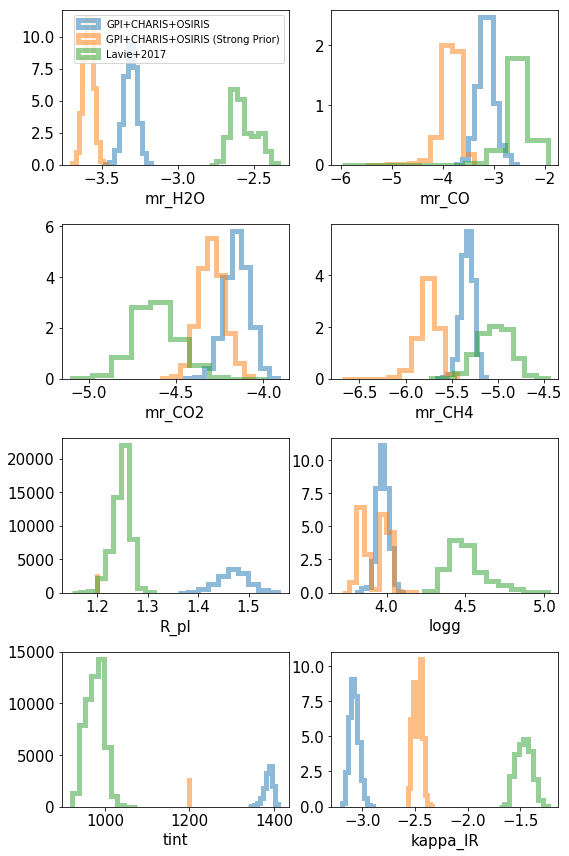}
\caption{probability density function for common parameters between ~\citet[green, ][]{Lavie2017} and our work. Blue is for the weak-prior case (\S \ref{sec:weak_priors}) and orange is for the strong-prior case (\S \ref{sec:strong_priors}). See Table \ref{tab:prior} for definition of each parameter. 
\label{fig:comp_us_lavie}}
\end{figure}

We first compare retrieved parameters that are common in our analyses (Fig. \ref{fig:comp_us_lavie}). For plotting clarity, we omit the result of adding $L$- and $M$-band data because the result is bracketed by the weak- and strong-prior cases. 

H$_2$O (major O carrier) and CO abundances (major C carrier) from ~\citet{Lavie2017} are higher, explaining the super-stellar C and O abundances in Fig. \ref{fig:co}. Abundances for CO$_2$ and CH$_4$ are low in all analyses, and therefore do not contribute significantly in C and O abundance. We will explain in \S \ref{sec:co2_ch4_det} why the abundances for CO$_2$ and CH$_4$ are detectable despite being relatively low. 

Planet radius and surface gravity retrieved by~\citet{Lavie2017} are $\sim$1.25 R$_{\rm{Jupiter}}$ and 4.5, which implies the inferred planet mass is 20 M$_{\rm{Jupiter}}$, too massive from a dynamical stability point of view~\citep{Fabrycky2010, Wang2018b}. As we discussed in \S \ref{sec:weak_priors}, strong priors need to be put on planet radius in order to satisfy the dynamical stability criterion. 

\subsubsection{P-T Profile}

The opacity retrieved in our work differs from that of~\citet{Lavie2017} by at least one order of magnitude. However, we used different formula for P-T profile calculation. We used an analytical function from~\citep{Parmentier2015} and ~\citet{Lavie2017} used a reduced version of equation 126 in~\citet{Heng2014}. A more direct comparison would be the P-T profiles that are calculated by different methods. 

\begin{figure}[h!]
\epsscale{1.0}
\plotone{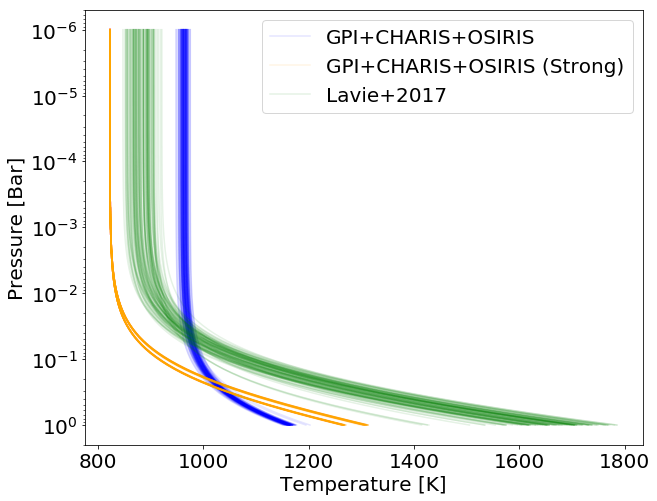}
\caption{Pressure-Temperature (P-T) profiles that are drawn from posterior distributions. Blue profiles are for the weak-prior case (\S \ref{sec:weak_priors}), orange profiles are for the strong-prior case (\S \ref{sec:strong_priors}), and green profiles are from~\citet{Lavie2017}.   
\label{fig:p_t}}
\end{figure} 

Fig. \ref{fig:p_t} shows P-T profiles using parameters that drawn from posterior distributions. Our results with weak priors shows highest temperature at low pressure levels but increases slowly toward higher pressures. This is due to the small retrieved opacity that causes a less sensitive temperature to pressure. The overall shape of P-T profile is similar between~\citet{Lavie2017} and our results with strong priors, but with temperature offset. Our P-T profile is lower by $\sim$100 K than that from~\citet{Lavie2017} although our retrieved t$_{\rm{int}}$ is $\sim$200 K hotter than theirs. 

\subsubsection{Cloud}

We assume global cloud that is parameterized by cloud top pressure, a cloud treatment from~\petit. At a first glance, the treatment is different from~\citet{Lavie2017} where three parameters are used to describe the global cloud: particle size, extinction efficiency, and cloud mixing ratio. In the case of planet c, the result from~\citet{Lavie2017} retrieved a cloud with refractory species (e.g., silicates) with particle size ($\sim$30 $\mu$m) larger than observed wavelength ($<$2.5 $\mu$m), which suggests a wavelength-independent cloud opacity in the near infrared. This is consistent with our cloud treatment with single parameter. One caveat though is that it is not entirely clear how to quantitatively compare our one-parameter model to their three-parameter model. We retrieved a cloud-top pressure of 1-2 bar, which indicates the global cloud is quite deep compared to that on hot Jupiters. 

\subsection{Comparing to~\citet{Konopacky2013}}
~\citet{Konopacky2013} conducted pioneering work on measuring chemical abundances of HR 8799 c. Unlike our free retrieval approach, they used a forward modeling approach, in which physically and chemically motivated models are used to fit the data. The two approaches may result in the difference of abundance measurements as shown in Fig. \ref{fig:co}. One outstanding issue is the non-detection of CH$_4$ and CO$_2$ in the original analysis in~\citet{Konopacky2013}. Below we offer some insights as to why CH$_4$ and CO$_2$ can be constrained by OSIRIS data alone. Their abundance can be further constrained by adding additional data as shown in~\citet{Lavie2017} and this work.  

\subsubsection{{{Constraining CO$_2$ and CH$_4$ Abundances?}}}
\label{sec:co2_ch4_det}

\begin{figure}[h!]
\epsscale{1.0}
\plotone{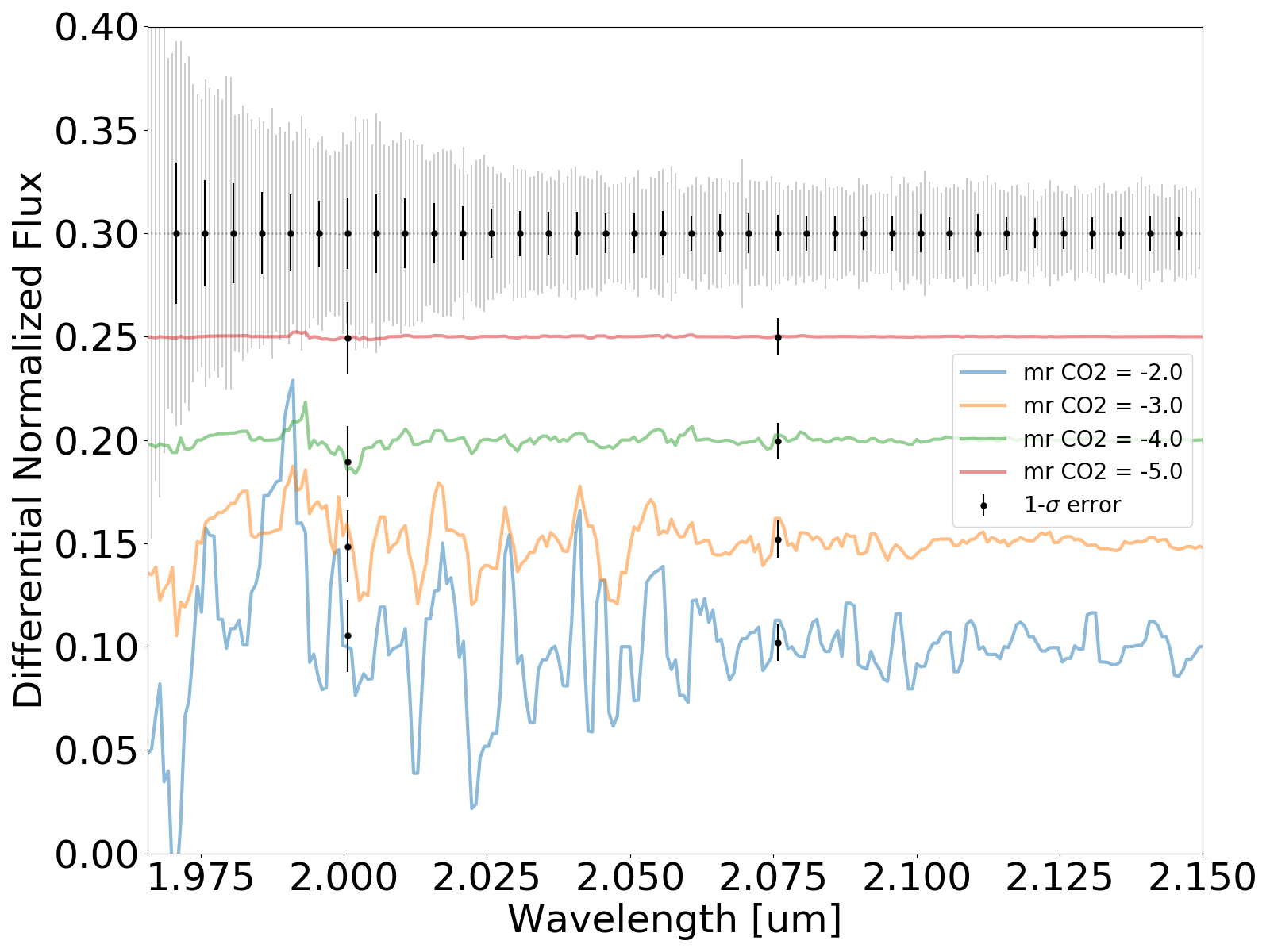}
\caption{Differential normalized flux for CO$_2$. We use \petit\ to generate spectra by varying only CO$_2$ abundance. Then, we normalize the spectra and subtract the normalized spectrum with zero CO$_2$ abundance. This results in the plotted differential normalized flux. For comparison, error bars reported in~\citet{Konopacky2013} are shown in grey and 1-$\sigma$ errors that correspond to the width of CO$_2$ spectral features are plotted in black. {{When log(mr$_{\rm{CO_2}}$) is between -3 (orange) and -4 (green), which is the range of retrieved CO$_2$ abundance (Table \ref{tab:mcmc_result}), CO$_2$ spectral features are comparable to 1-$\sigma$ errors, suggesting that it is possible to constrain CO$_2$ abundance with the OSIRIS data quality. }}
\label{fig:CO2_signal_K}}
\end{figure}
\begin{figure}[h!]
\epsscale{1.0}
\plotone{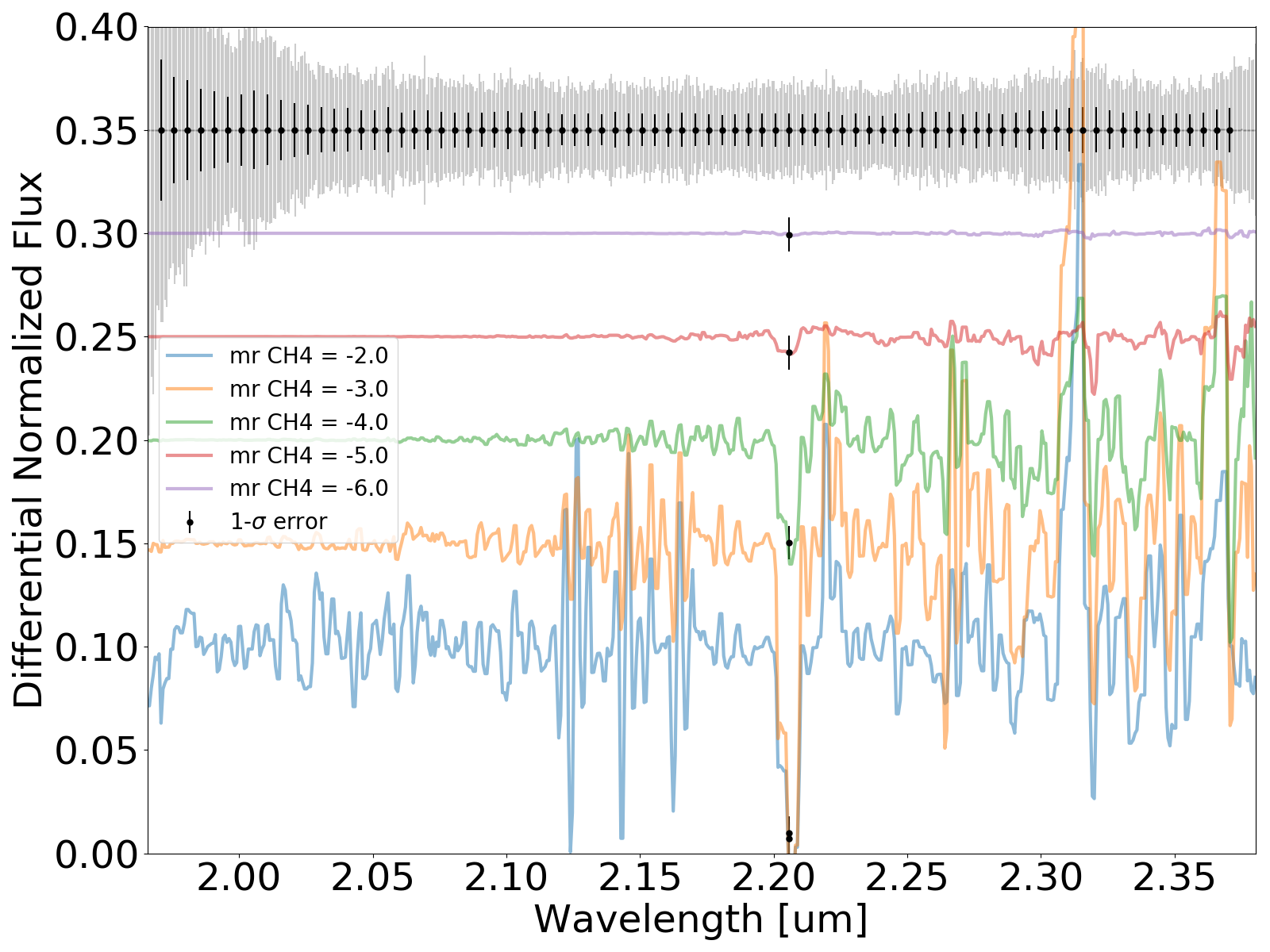}
\caption{Same as Fig. \ref{fig:CO2_signal_K} except for CH$_4$. {{When log(mr$_{\rm{CH_4}}$) is between -4 (green) and -5 (red), which is the range of retrieved CH$_4$ abundance (Table \ref{tab:mcmc_result}), CH$_4$ spectral features are comparable to 1-$\sigma$ errors, suggesting that it is possible to constrain CH$_4$ abundance with the OSIRIS data quality. }}
\label{fig:CH4_signal_K}}
\end{figure}

{While our results indicate that CO$_2$ and CH$_4$ abundances are constrained, we have the following reasons to cast a doubt on the constrain. 

First, CO$_2$ abundance has never been constrained in previous retrieval works on brown dwarfs~\citep{Line2015, Madhusudhan2016,Line2017,Zalesky2019,Kitzmann2020}. Indeed, CO$_2$ abundance is expected to be $\sim$4 orders of magnitude lower than CO for a brown dwarf with similar effective temperature~\citep{Sorahana2012}. This is at odds with our CO$_2$ values. 

Second, including CO$_2$ and CH$_4$ in the retrieval could further increase the likelihood because changing CO$_2$ and CH$_4$ abundance may look like a change in the continuum shape at the OSIRIS S/N level, therefore these extra free parameters help to fit the data better. This argument may be supported by that GPI data alone can constrain CO$_2$ abundance (\S \ref{sec:gpi_result}).   

Third, we investigate if the presence of CO$_2$ or CH$_4$ can be visually discerned at the level of log(mr$_{\rm{CO_2}}$) between -3 and -4 and log(mr$_{\rm{CH_4}}$) between -4 and -5. For reference the retrieved log(mr$_{\rm{CO_2}}$) and log(mr$_{\rm{CH_4}}$) is around $\sim$-3 and between -4.5 and -5.0 (see Table \ref{tab:mcmc_result}). We find that the OSIRIS 1-$\sigma$ error bars are comparable with spectral features of CO$_2$ or CH$_4$. Therefore, we cannot confidently claim detection of CO$_2$ or CH$_4$. 

Fig. \ref{fig:CO2_signal_K} and Fig. \ref{fig:CH4_signal_K} show the difference spectra at different mixing ratio of CO$_2$ or CH$_4$. The difference spectra are created by subtracting two spectra: one with a specified mixing ratio and the other with a zero mixing ratio for a given species. The difference spectra, when compared to error bars (also shown in Fig. \ref{fig:CO2_signal_K} and Fig. \ref{fig:CH4_signal_K}), would suggest if the data have the distinguishing power for a give mixing ratio. Indeed, Fig. \ref{fig:CO2_signal_K} and Fig. \ref{fig:CH4_signal_K} show that spectral features of CO$_2$ and CH$_4$ at mixing ratios of log(mr$_{\rm{CO_2}}$) between -3 and -4 and log(mr$_{\rm{CH_4}}$) between -4 and -5 are larger or comparable with error bars, and therefore can be distinguished by the data. }

{
\subsubsection{The Contribution of CO$_2$ and CH$_4$ Abundances to C/O}
\label{sec:co2_to_co}
The abundance of CH$_4$, even in the constrained case, is at least two orders of magnitude lower than CO abundance. Therefore, CH$_4$ is not the major carbon carrier and does not significantly change the total carbon abundance in the C/O calculation. 

{Whether or not constraining/including CO$_2$ abundance could significantly change the measured C/O. We have run a retrieval analysis on the full data sets (including $L$- and $M$-band data) assuming no CO$_2$ and CH$_4$, The resulting [C/H] and [O/H] deviates by $\sim$2-$\sigma$ and C/O by $\sim$1-$\sigma$ from the case that includes CO$_2$ and CH$_4$ (Table \ref{tab:mcmc_result} and Fig. \ref{fig:co}). 

While Bayesian evidence strongly favors the model that includes CO$_2$ and CH$_4$ ($\Delta\ln(Z)=22.5$), the evidence laid out in \S \ref{sec:co2_ch4_det} casts a reasonable doubt on the preferred model by the Bayesian evidence. Additionally, we note that our models are not a fully accurate representation of the true spectrum. Favoring a more complex model with nonphysical parameters could indicate that there are deficiencies in our models that the CH$_4$ and CO$_2$ parameters are trying to account for. Moreover, systematics in the data that are not accounted for by the error bars or by uncorrelated noise could introduce spurious detection. }




}

\subsection{Retrieval on Individual Data Set}

We perform retrieval on individual data set from GPI, OSIRIS, and CHARIS. The results are provided in \S \ref{app:individual_analysis}. Here we summarized the major findings and the lesson learned. 

We propose two ways of comparing retrieval results. The first way is to compare C/O. We show that C/O ratios retrieved from individual data set are consistent except for CHARIS retrieval because CHARIS data do not constrain the major carbon carrier CO (Fig. \ref{fig:comp_co_individual}). In principle, we can compare C/H and O/H, but the normalization process of the OSIRIS data makes it difficult to do so. 

Secondly, comparing the likelihood distribution can check consistency of retrieval results. We show that GPI $H$ and $K$ band data return consistent retrieval results (Fig. \ref{fig:gpi_h_k_comp}). This demonstrates the internal consistency of GPI data. In contrast, we show that retrieval results between GPI and OSIRIS are inconsistent by comparing the distribution of likelihood values (Fig. \ref{fig:prob_comp_gpi_osiris}). The most likely explanation is the different systematics that exist in individual data set. 

To summarize, combining data sets that are taken using different instruments at different epochs comes with a caveat: astrophysical conditions and instrument-induced systematics may play a critical role in causing inconsistent retrieval results. For future observations, it is ideal to (1) obtain simultaneous data that cover wide wavelengths and (2) understand better systematics that is caused by different instruments. 

\subsection{$\lambda$ Bootis Nature of HR~8799}
\label{sec:lambdaBootis}

{{ \citet{Gray1999} and \citet{Sadakane2006} have confirmed that HR~8799 is a mild $\lambda$~Bootis star. Concerning the origin of the abundance anomalies detected in $\lambda$~Bootis type stars, several scenarios have been proposed. One scenario suggests that at least a part of $\lambda$~Bootis type stars can be explained by assuming a binary system consisting of two stars of similar spectral types \citep{Andrievsky1997, Faraggiana1999, Gerbaldi2003}. Since \citet{Mathias2004} and \citet{Henry2005} concluded that HR~8799 is a single star based on radial velocity observations, the second scenario is ruled out.

Another widely accepted scenario invokes the process of selective accretion of circum-stellar or inter-stellar material~\citep[ISM,][]{Venn1990, Turcotte1993, Gray2002} by a solar abundance star. For example,~\citet{Su2009} studied the environment of HR~8799 using infrared imaging data from {\it Spitzer} and CO(3-2) map from the James Clerk Maxwell Telescope (JCMT). They detected a molecular cloud that is possibly associated with HR~8799, and suggested accretion from this cloud might explain the abundance anomalies of HR~8799. 

\citet{Jura2015} argued that the accretion rates of ISM required for the occurrence of $\lambda$ Bootis phenomena are unachievable, and instead suggested that accretion of winds from an additional hot Jupiter companion might account for the $\lambda$~Bootis nature of HR~8799. This idea has support from observations that A-type stars are efficient at forming giant planets~\citep{Johnson2011, Murphy2016}, and some of these giant planets are found in tight orbits around the star~\citep{Collier2010}. However, detection of such an unknown planet around HR~8799 is difficult using current observing techniques.

\citet{Draper2016} studied a sample of $\lambda$ Bootis stars using {\it Herschel Space Observatory}, and found IR excesses around nearby $\lambda$ Boo stars are caused by debris disks rather than ISM bow waves. Indeed, HR~8799 has a resolved debris disk based on {\it Herschel} data~\citep{Matthews2014}. Since a higher influx of planetesimals and comets could provide enough volatile gas for accretion, the planet systems in HR~8799 could possibly trigger higher than usual dynamic activities in the disk, which provide accretion material to the star and cause the $\lambda$ Bootis nature of HR~8799. Currently this is the most probably scenario. However, further theoretical study on the transport of circum-stellar material and stellar mixing mechanisms are required to establish a relation between the stellar surface abundance anomalies and external accretion process. 

Contrary to the above debris-disk theory, \citet{Moya2010} studied the $\lambda$ Bootis nature of HR~8799 using a comprehensive seismology modeling. They found the observation data contradicts one of the main hypotheses for explaining the $\lambda$ Bootis nature, namely the accretion/diffusion of gas by a star with solar abundance. However, their results are dependent on the inclination angles of HR~8799 and the correct mode identification, which are vital for the asterseismic analysis. Thus, more observational and theoretical investigations are still needed to pin down the origin of the $\lambda$ Bootis nature of HR~8799. 

 }}

\section{Summary}
\label{sec:summary}

\begin{itemize}
    \item We measure elemental abundance for HR 8799 by analyzing data taken from LBT/PEPSI and ESO-3.6m/HARPS using spectral analysis package MOOG. We report stellar [C/H], [O/H], and C/O to be 0.11$\pm$0.12, 0.12$\pm$0.14, and 0.54$^{+0.12}_{-0.09}$. The result is summarized in Table \ref{tab:CO_abundances}. 
    \item We develop a retrieval code that uses \petit\ to model planet atmospheres and \pmn\ to sample posterior distributions. Based on the combined data sets from Gemini/GPI, Keck/OSIRIS, and Subaru/CHARIS, we measure planet [C/H], [O/H], and C/O for HR 8799 c to be 0.16$^{+0.12}_{-0.13}$, 0.13$^{+0.08}_{-0.08}$, and 0.58$^{+0.06}_{-0.06}$.  
    \item However, the above retrieval results lead to a planet that may be too massive to maintain dynamical stability. After applying a strong prior to ensure dynamical stability, the planet [C/H] and [O/H] change by 0.57 and 0.37 dex. we measure planet [C/H], [O/H], and C/O for HR 8799 c to be -0.41$^{+0.11}_{-0.12}$, -0.26$^{+0.05}_{-0.06}$, and 0.39$^{+0.06}_{-0.06}$. 
    \item {After adding $L$- and $K$-band photometric data points, the retrieval result is closer to that of a more physically and chemically motivated forward modeling approach~\citep{Konopacky2013}. This suggests that adding data points at longer wavelengths helps to constrain the spectral energy distribution and thus the abundances and cloud condition. However, discrepancy remains, and the level of discrepancy depends on if CO$_2$ and CH$_4$ abundances are actually constrained. This is discussed in \S \ref{sec:co2_ch4_det} and \S \ref{sec:co2_to_co}. }
    \item A summary of the retrieval results can be found in Table \ref{tab:CO_abundances} and \ref{tab:mcmc_result} and Fig. \ref{fig:co}. The implications are discussed in \S \ref{sec:co}. 
    \item When compared to previous results, we find the abundances varies by more than 1 dex, highlighting the uncertainty in atmospheric retrieval. We discuss potential issues that could lead to the differences in \S \ref{sec:dis} and \S \ref{app:individual_analysis}.
    \item The inconsistency in retrieval results could stem from different treatments in modeling planet atmospheres, e.g., P-T profile calculation and cloud modeling, or the different data sets that used in retrieval. We show in \S \ref{app:individual_analysis} that systematics in individual data set, be it astrophysical or instrumental, can results in significantly different results. {{We therefore advocate for (1) providing details for modeling planet atmospheres to facilitate comparison with other investigations; (2) simultaneous observations that span a wide wavelength range to minimize the heterogeneous contribution from temporal astrophysical variations and/or systematics from using different data sets; and (3) a better understanding of systematic errors that are introduced by different instruments. }}
\end{itemize}

\noindent
{\bf{Acknowledgments}} We would like to thank Quinn Konopacky for providing Keck/OSIRIS data, Baptieste Lavie for providing the retrieval posterior samples for comparison, Jonathan Fortney for insightful discussions on atmospheric retrieval, Paul Molliere for the help in setting up and running \petit. We thank the anonymous referee for his/her constructive comments and suggestions that greatly improve the manuscript. We thank the Heising-Simons Foundation for supporting the workshop on combining high-resolution spectroscopy and high-contrast imaging for exoplanet characterization, where the idea originated on combining photometric data and spectral data of different resolutions. BM thank the support of CSST grant and NSFC grant U1931102. The data presented herein were obtained at the W. M. Keck Observatory, which is operated as a scientific partnership among the California Institute of Technology, the University of California and the National Aeronautics and Space Administration. The Observatory was made possible by the generous financial support of the W. M. Keck Foundation. The authors wish to recognize and acknowledge the very significant cultural role and reverence that the summit of Mauna Kea has always had within the indigenous Hawaiian community.  We are most fortunate to have the opportunity to conduct observations from this mountain. {{We are grateful for the PEPSI and LBTI data from LBT. The LBT is an international collaboration among institutions in the United States, Italy and Germany. The LBT Corporation partners are: The University of Arizona on behalf of the Arizona university system; Istituto Nazionale di Astrofisica, Italy;  LBT Beteiligungsgesellschaft, Germany, representing the Max Planck Society, the Astrophysical Institute Potsdam, and Heidelberg University; The Ohio State University; The Research Corporation, on behalf of The University of Notre Dame, University of Minnesota and University of Virginia. We thank the LBTI team for offering their instrument to the community and the opportunity to attempt the parallel observations with PEPSI. Jordan Stone is supported by NASA through Hubble Fellowship grant HST-HF2-51398.001-A awarded by the Space Telescope Science Institute, which is operated by the Association of Universities for Research in Astronomy, Inc., for NASA, under contract NAS5-26555. }}

%






\appendix
\section{Retrieval results for individual dataset}
\label{app:individual_analysis}

In addition to analyzing the data set that combines OSIRIS, GPI, and CHARIS data, we analyze each component of the data set individually. The retrieval results and our notes are shown in this section and Table \ref{tab:mcmc_result}.

\subsection{Keck/OSIRIS}

\subsubsection{Absolute Flux}
\label{sec:osiris_result}
\begin{figure}[h!]
\epsscale{1.0}
\plotone{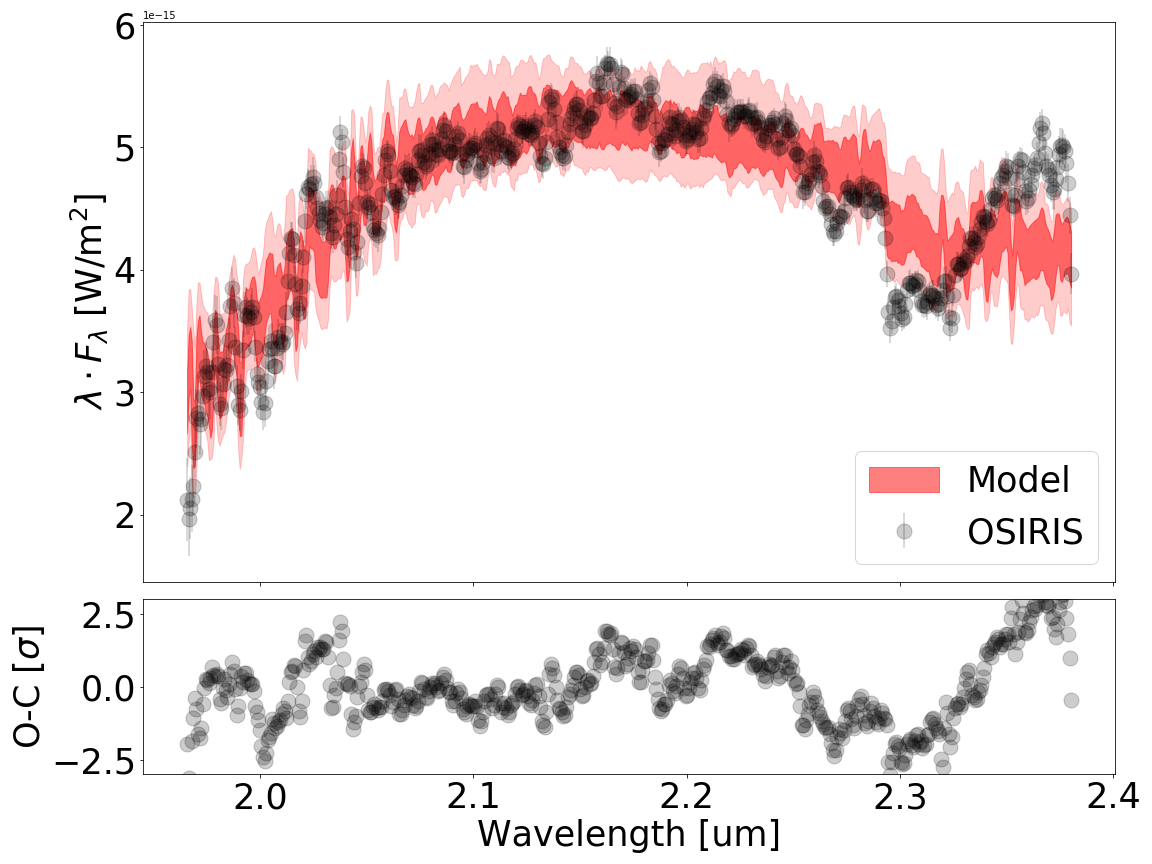}
\caption{Top: OSIRIS data points (black) and modeled spectra with retrieved parameters (red). Dark and light shaded regions represent 68\% and 95\% credible intervals. Bottom: residual plot with data minus model and divide by errors. Errors are the quadrature summation of OSIRIS reported error bars and the 1-$\sigma$ dispersion of the modeled spectra.
\label{fig:osiris_abs_data_model}}
\end{figure} 
We cannot obtain a reasonable agreement between our modeled spectra and the absolute flux measured from OSIRIS. This is evident from the residual plot in Fig. \ref{fig:osiris_abs_data_model}. The residual has a clear correlated pattern especially towards the longer wavelengths beyond 2.3 $\mu$m. We attribute the uncounted residual pattern to the overall spectral energy distribution and the longer wavelength end of the OSIRIS data. More specifically, the OSIRIS data is more rounded than the model and the OSIRIS flux between 2.3 and 2.4 $\mu$m increases whereas the model flux flattens in the same wavelength region. We suspect that OSIRIS data have some systematic error that alters its spectral shape, which results in the challenges in our retrieval using OSIRIS absolute flux. In principle, the systematics can be taken care of by the Gaussian process. However, the Gaussian process may not be the best tool to model the two aforementioned spectral morphological differences. 

\subsubsection{Normalized Flux}

\begin{figure}[h!]
\epsscale{1.0}
\plotone{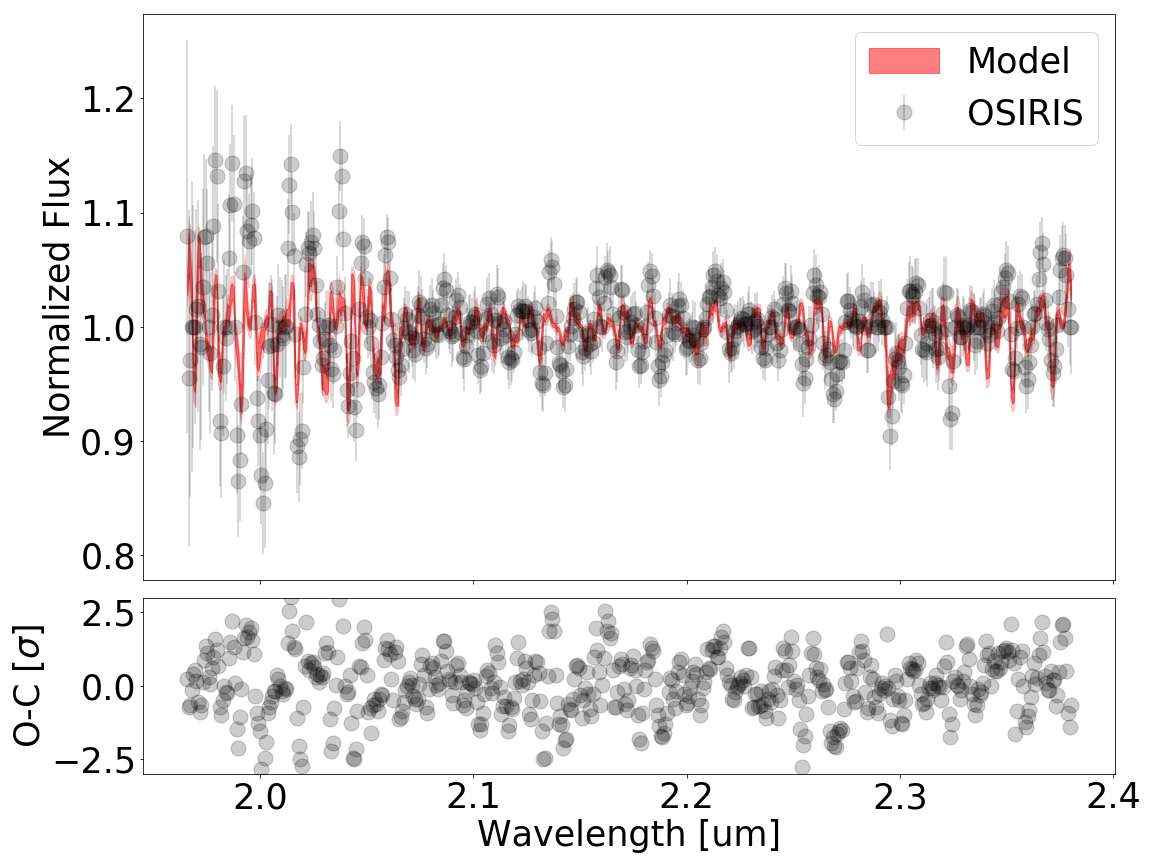}
\caption{Same as Fig. \ref{fig:osiris_abs_data_model} but for normalized OSIRIS data with weak priors as shown in Table \ref{tab:prior}.  
\label{fig:osiris_norm_weak_data_model}}
\end{figure} 

\begin{figure}[h!]
\epsscale{1.0}
\plotone{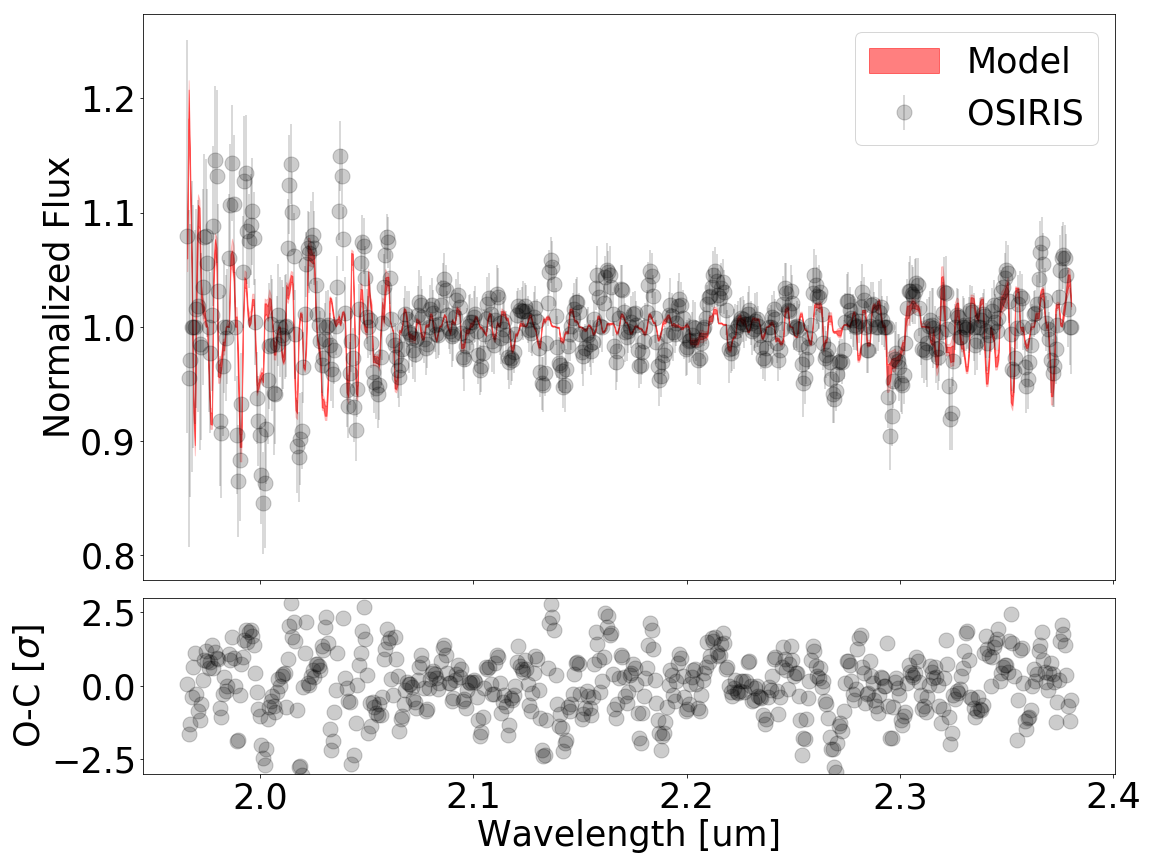}
\caption{Same as Fig. \ref{fig:osiris_abs_data_model} but for normalized OSIRIS data with strong priors.  
\label{fig:osiris_norm_strong_data_model}}
\end{figure} 
If systematics only affects low-frequency part of the spectrum, e.g., overall spectral shape etc., then we can remove the low-frequency systematics with a high-pass filter and use the normalized flux for retrieval.

After normalization, the continuum information is missing. This causes the negative correlation between $\kappa_{\rm{IR}}$ vs. P$_{\rm{cloud}}$. The two parameters compensated each other: the same observed line depth can be achieved by either a high $\kappa_{\rm{IR}}$ or a high P$_{\rm{cloud}}$ (i.e., low or no cloud). However, the degeneracy of the two parameters cannot be broken by the normalized-flux data set. In addition, the retrieved mixing ratios are too high than physically allowed especially for CO and H$_2$), e.g., log(mr$_{\rm{H_2O}}$) = -0.81 and log(mr$_{\rm{CO}}$)= -0.61. 

To alleviate the above issues, we fix the following values in retrieval because they are more physically plausible and constrained by other date sets whose continuum information is not lost due to the normalization. The parameters that we fixed are: $\log$g = 3.95, R$_P$ = 0.90 R$_{\rm{Jupiter}}$, log(P$_{\rm{cloud}}$) = -0.05, log($\kappa_{\rm{IR}}$) = -2.11, and t$_{\rm{int}}$ = 1213 K. The retrieved mixing ratios are more reasonable when compared to retrieval results using other data sets. However, the log likelihood drops by $\sim$25, which is also evident from the slight more scatter when comparing the residual plots between Fig. \ref{fig:osiris_norm_weak_data_model} and Fig. \ref{fig:osiris_norm_strong_data_model}. This highlights the limitation of retrieval: the most likely parameter space to fit the data may not be physically/chemically allowed.

\subsection{Gemini-S/GPI}
\label{sec:gpi_result}
\begin{figure}[h!]
\epsscale{1.0}
\plotone{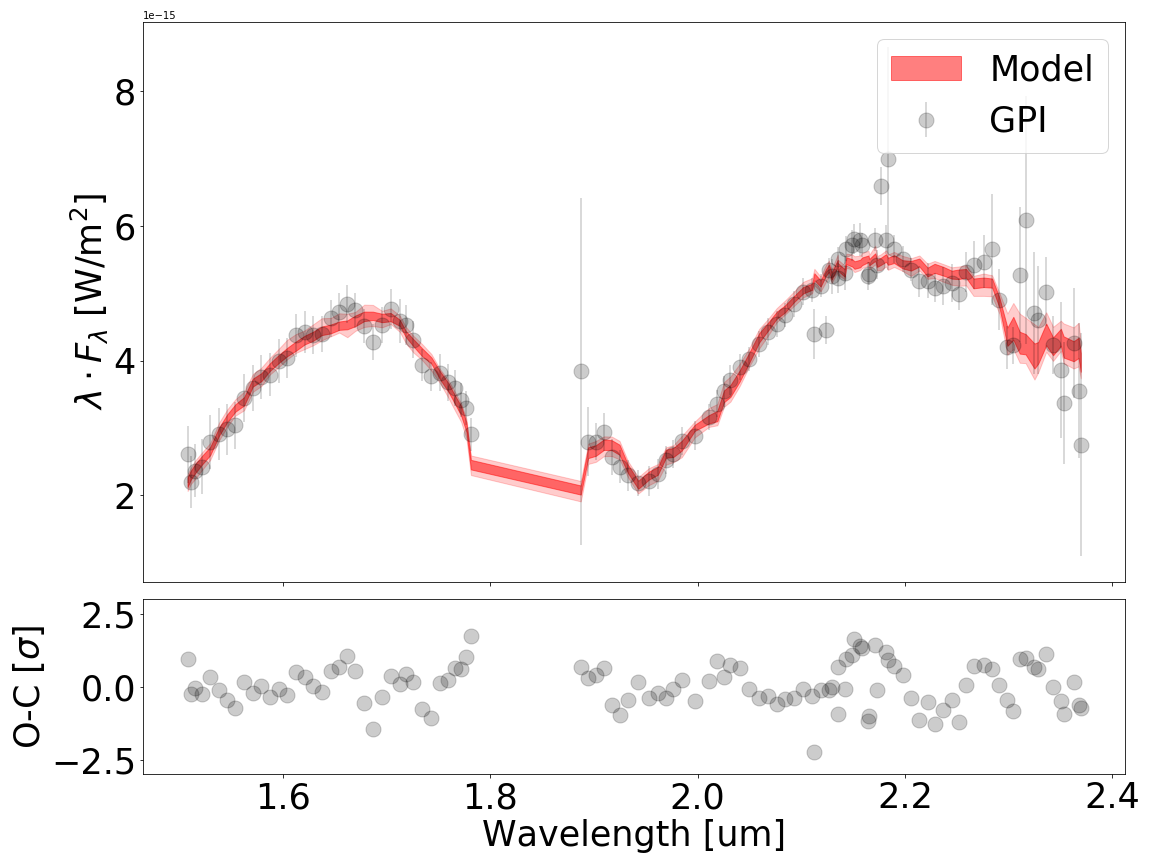}
\caption{Same as Fig. \ref{fig:osiris_abs_data_model} but for GPI data.  
\label{fig:gpi_data_model}}
\end{figure} 


Spectra based on GPI retrieval results are shown in Fig. \ref{fig:gpi_data_model}. Modeled spectra trace data points reasonably well in both $H$ and $K$ band. All parameters are well constrained except for CH$_4$. 

\subsection{Subaru/CHARIS}
\label{sec:charis_result}
\begin{figure}[h!]
\epsscale{1.0}
\plotone{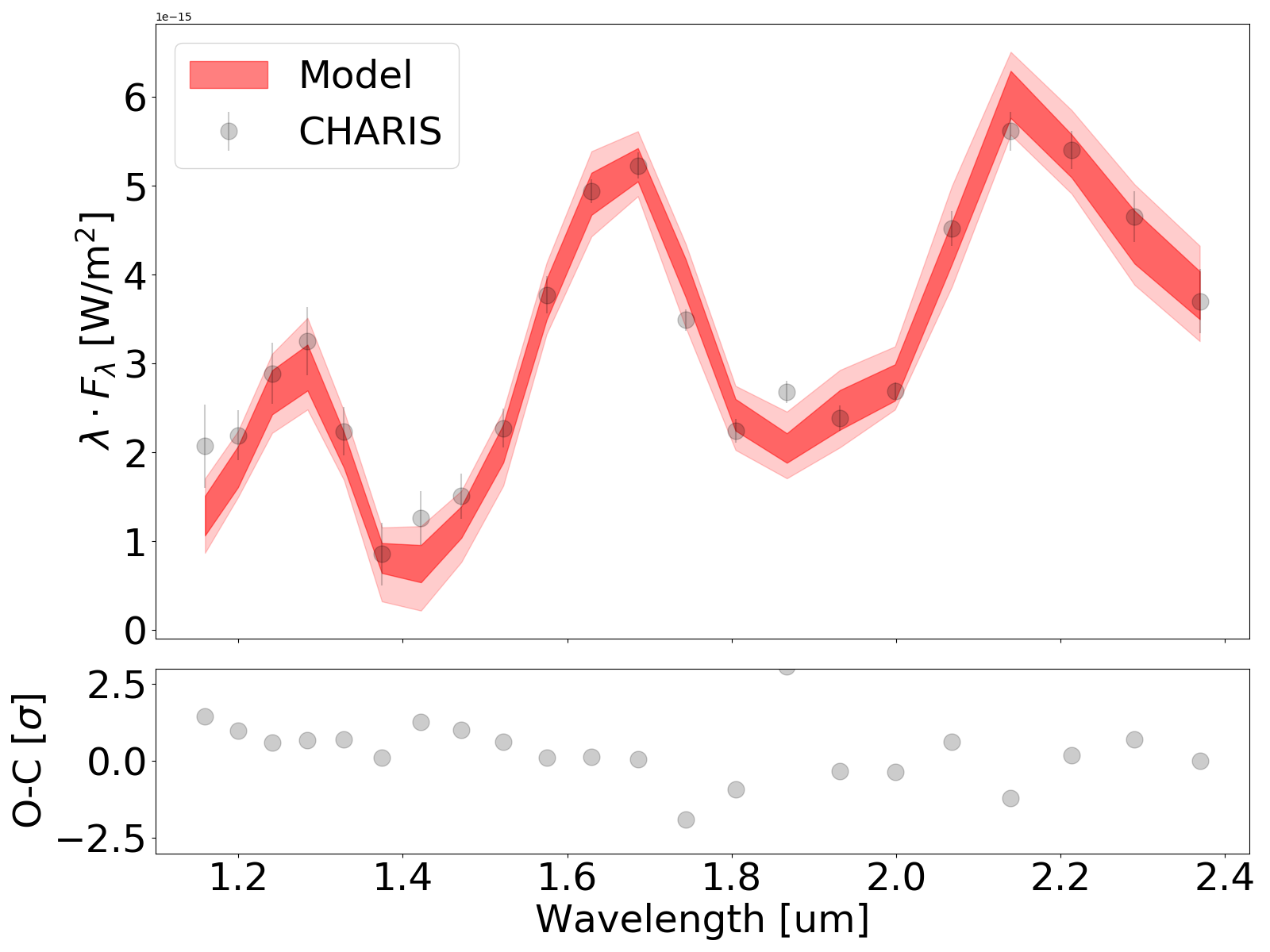}
\caption{Same as Fig. \ref{fig:osiris_abs_data_model} but for CHARIS data.  
\label{fig:charis_data_model}}
\end{figure} 

Modeled spectra based on CHARIS retrieval are shown in Fig. \ref{fig:charis_data_model}. The modeled spectra generally agree with data points, but the number of data points are sparse due to the low spectral resolution. 

\section{Comparing Results From Individual Datesets}
\label{sec:quant_comp}
In order to compare retrieval results from different data sets, we need to define a set of criteria for the comparison. by ``compare", we mean to check the consistency of results. 
\subsection{C/O}
\begin{figure}[h!]
\epsscale{1.0}
\plotone{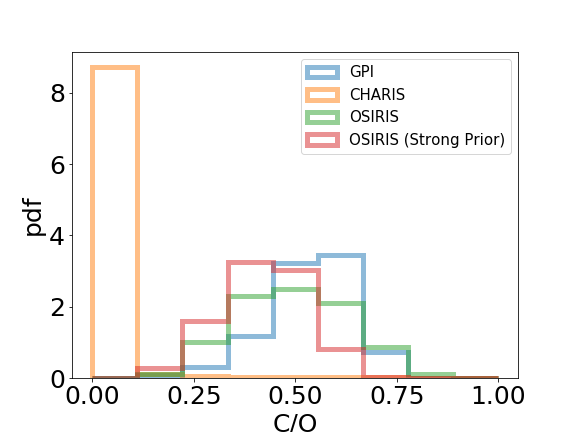}
\caption{Comparing C/O posterior distributions using individual data sets.  
\label{fig:comp_co_individual}}
\end{figure}

Because we have normalized data from OSIRIS, which are insensitive to absolute molecular abundances, we only compare C/O that are retrieved (Fig. \ref{fig:comp_co_individual}). GPI and OSIRIS results are generally consistent. However, the result from the normalized OSIRIS data with strong priors skews C/O to lower values, in a similar way that the C/O is lower in the strong-prior case for the combined data (\S \ref{sec:strong_priors}). C/O from the CHARIS data peak at zero because H$_2$O is constrained but the major carbon carrier CO is unconstrained.  

\subsection{Likelihood}

Comparing likelihood is another way of checking consistency between retrieval results. More specifically, Two likelihoods will be calculated and compared. The first likelihood is the likelihood of one retrieval as calculated from the posteriors. The second likelihood is the likelihood when plugging posteriors of another retrieval. If two results are consistent, then the two likelihoods should share similar statistics (e.g., median or mode) or follow the same distribution. Otherwise, the two retrieval results are inconsistent with each other. 

\subsubsection{GPI $H$ vs. $K$ band}

We divide GPI data into two parts, $H$ and $K$ band, and conduct atmospheric retrieval for the two data sets. This is to check the internal consistency of GPI data. If retrieval results are similar between $H$ and $K$ band, i.e., retrieved parameters share similar posterior distributions and similar posterior values when plugging posteriors from $H$ band retrieval and when plugging posteriors from $K$ band retrieval. 

\begin{figure*}
  \centering
  \begin{tabular}[b]{cc}
    \includegraphics[width=.45\linewidth]{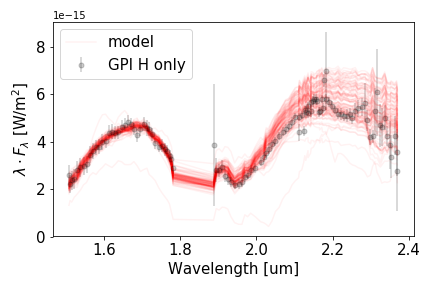} & \includegraphics[width=.45\linewidth]{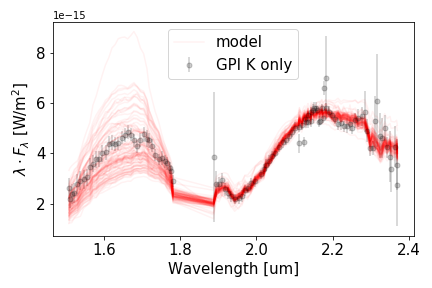} \\
    \includegraphics[width=.45\linewidth]{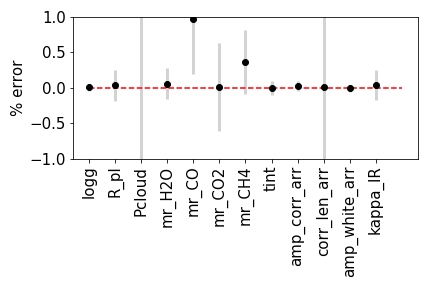} & \includegraphics[width=.45\linewidth]{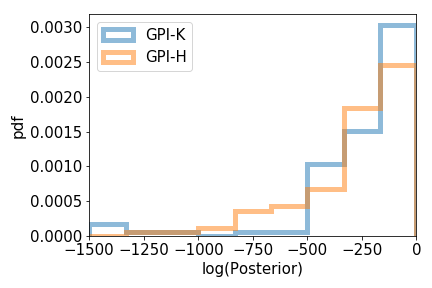} \\

  \end{tabular} \qquad
  \caption{Top left: Modeled spectra drawn from posterior distribution (red, based on $H$ band data) are compared with the GPI IFS data. Top right: Modeled spectra drawn from posterior distribution (red, based on $K$ band data) are compared with the GPI IFS data. Bottom left: {{comparing retrieved parameters from the $H$-band only data and the $K$-band only data. The retrieved parameters are generally consistent with each other. The inconsistency in CO abundance is due to the lack of constraining power on CO abundance in $H$-band only data}}. Bottom right: the distribution of posterior values for $K$-only retrieval (blue) and the distribution of posterior values for $H$-only retrieval (orange). {{The two posterior distributions are indistinguishable (p=0.15 from a two-sample K-S test) suggests that the results are consistent between $H$-only retrieval and $K$-only retrieval. }}  \label{fig:gpi_h_k_comp}}
\end{figure*} 

Fig. \ref{fig:gpi_h_k_comp} shows the results. Top two panels show models with parameters drawn from posterior distributions from $H$ and $K$ band data. It is understandable that models using retrieval results of $H$ (or $K$) band only agree well with $H$ (or $K$) band data. 

Retrieved parameters share similar values as shown on the lower left panel in Fig. \ref{fig:gpi_h_k_comp}. We use percentage error here with zero as zero percent and unity as 100\%. The percentage error is calculated as the difference of median between $H$ and $K$ band results divided by the average of the medians. The error associated with each point is calculated as follows: we add in quadrature the average of reported upper and lower error bars for each band, and divide the value by the average the medians. The majority of parameters as retrieved from $H$ and $K$ band agree well with a few percent except for P$\rm{cloud}$, $\log$(mr$_{\rm{C}\rm{O}}$), and $\log$(mr$_{\rm{C}\rm{H}_4}$). P$\rm{cloud}$ has a value that is very close to zero, and therefore the large percentage error and the error bar. H band data do not provide a strong constraint on $\log$(mr$_{\rm{C}\rm{O}}$) and $\log$(mr$_{\rm{C}\rm{H}_4}$) because of weak absorption of these two species and therefore the disagreement between $H$ and $K$ band results. Overall, $H$ and $K$ band results are consistent within a few percent.

The consistency between $H$ and $K$ band results is further supported by the comparison of posterior values (lower right panel in Fig. \ref{fig:gpi_h_k_comp}). Posterior value distribution from $H$-band retrieval is in good agreement with the posterior value distribution using retrieval results from $K$ band (two-sample K-S test p = 0.15). 



\subsubsection{GPI vs. OSIRIS}
\begin{figure}
  \centering
  \begin{tabular}[b]{c}
    \includegraphics[width=.5\linewidth]{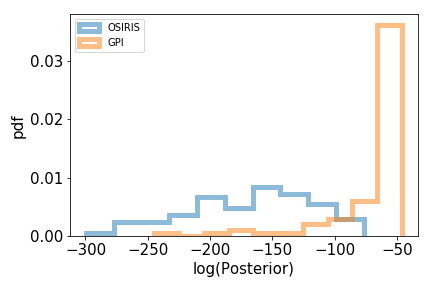}  \\
  \end{tabular} \qquad
  \caption{Comparing the distribution of posterior values using GPI retrieved parameters (orange) and posteior values using OSIRIS retrieved parameters (blue). \label{fig:prob_comp_gpi_osiris}}
\end{figure}

Comparing GPI and OSIRIS retrieval results is more complicated because the two data sets are taken by different instruments and at different times. Astrophysical time variability may be one of the reasons that causes inconsistency, if any. Among parameters that are used in the retrieval process, P$_{\rm{cloud}}$ may be variable over time. In addition, we used the same $\log g$, R$_{\rm{pl}}$, and t$_{\rm{int}}$ that are retrieved from GPI in OSIRIS retrieval. Therefore in the following likelihood comparison, we only plug in posteriors of chemical abundance from OSIRIS into GPI likelihood function. Fig. \ref{fig:prob_comp_gpi_osiris} shows that the likelihood distribution already looks significantly different with a K-S test p value of $2.9\times10^{-32}$. The mode of the OSIRIS likelihood distribution is at the 1 percentile of the GPI likelihood distribution. The retrieval results are inconsistent. The inconsistency suggests that astrophysical time variability is not the sole explanation because the potential time variable has been excluded in the comparison. Instrument- or date-reduction-induced systematics should be responsible for the inconsistency as observed in the retrieval results.

\bibliography{sample63}{}
\bibliographystyle{aasjournal}



\begin{deluxetable}{ccccc}
\tablewidth{0pt}
\tablecaption{Stellar Atmospheric Parameters of the star HR~8799 \label{tab:stellarparams}}
\tablehead{
\colhead{\textbf{$T_{\rm eff}$}} &
\colhead{\textbf{$\log(g)$}} &
\colhead{\textbf{[Fe/H]}} &
\colhead{\textbf{$\xi_{t}$}} &
\colhead{\textbf{Notes}} \\
\colhead{\textbf{K}} &
\colhead{\textbf{cgs}} &
\colhead{\textbf{dex}} &
\colhead{\textbf{km~s$^{-1}$}} &
\colhead{\textbf{}}
}


\startdata
7250 $\pm$ 150 & 4.30 $\pm$ 0.10 & -0.55 $\pm$ 0.11 & 3.3 $\pm$ 0.3  & PEPSI \\
7450 $\pm$ 100 & 4.40 $\pm$ 0.10 & -0.48 $\pm$ 0.12 & 2.9 $\pm$ 0.2  & HARPS \\
7390 $\pm$ 80 & 4.35 $\pm$ 0.07 & -0.52 $\pm$ 0.08 & 3.1 $\pm$ 0.2  & Combined 
\enddata

\end{deluxetable}

\begin{deluxetable}{clllll}
\tablewidth{0pt}
\tablecaption{C and O abundances for HR~8799 and planet c \label{tab:CO_abundances}}
\tablehead{
\colhead{\textbf{}} &
\colhead{\textbf{$\log{\epsilon_{\rm C}}$}} &
\colhead{\textbf{[C/H]}} &
\colhead{\textbf{$\log{\epsilon_{\rm O}}$}} &
\colhead{\textbf{[O/H]}} &
\colhead{\textbf{C/O}} 
}


\startdata
\multicolumn{6}{l}{HR 8799} \\
\hline
This work &8.54$\pm$0.12 & 0.11$\pm$0.12 & 8.81$\pm$0.14 & 0.12$\pm$0.14 & 0.54$^{+0.12}_{-0.09}$ \\
\citet{Sadakane2006} &8.63 & 0.20 & 8.88 & 0.19 & 0.56 \\
Solar &8.43 & 0.00 & 8.69 & 0.00 & 0.55 \\
\hline
\multicolumn{6}{l}{HR 8799 c} \\
\hline
Weak Priors &8.59$^{+0.12}_{-0.13}$ & 0.16$^{+0.12}_{-0.13}$ & 8.82$^{+0.08}_{-0.08}$ & 0.13$^{+0.08}_{-0.08}$ & 0.58$^{+0.06}_{-0.06}$ \\
Strong Priors &8.02$^{+0.11}_{-0.12}$ & -0.41$^{+0.11}_{-0.12}$ & 8.43$^{+0.05}_{-0.06}$ & -0.26$^{+0.05}_{-0.06}$ & 0.39$^{+0.06}_{-0.06}$ \\
L+M &8.31$^{+0.09}_{-0.10}$ & -0.11$^{+0.09}_{-0.10}$ & 8.61$^{+0.06}_{-0.06}$ & -0.07$^{+0.06}_{-0.06}$ & 0.49$^{+0.05}_{-0.04}$ \\
L+M (no CO$_2$) &8.13$^{+0.08}_{-0.10}$ & -0.29$^{+0.08}_{-0.10}$ & 8.50$^{+0.05}_{-0.05}$ & -0.18$^{+0.05}_{-0.05}$ & 0.43$^{+0.05}_{-0.05}$ \\
\citet{Lavie2017} &9.26$^{+0.15}_{-0.20}$ & 0.83$^{+0.15}_{-0.20}$ & 9.52$^{+0.09}_{-0.11}$ & 0.83$^{+0.09}_{-0.11}$ & 0.54$^{+0.11}_{-0.12}$ \\
\citet{Konopacky2013} &8.33$^{+0.02}_{-0.04}$ & -0.1$^{+0.02}_{-0.04}$ & 8.51$^{+0.06}_{-0.09}$ & -0.18$^{+0.06}_{-0.09}$ & [0.60..0.75] \\
\enddata

\end{deluxetable}

\begin{deluxetable}{lcccc}
\tablewidth{0pt}
\tablecaption{Parameters used in retrieval and their priors.\label{tab:prior}}
\tablehead{
\colhead{\textbf{Parameter}} &
\colhead{\textbf{Unit}} &
\colhead{\textbf{Type}} &
\colhead{\textbf{Lower}} &
\colhead{\textbf{Upper}} \\
\colhead{\textbf{}} &
\colhead{\textbf{}} &
\colhead{\textbf{}} &
\colhead{\textbf{or Mean}} &
\colhead{\textbf{or Std}} 
}


\startdata
Surface gravity ($\log$g)                   &  cgs                 &   Gaussian        &  4.0        &   0.1       \\
Planet radius (R$_P$)                       &  R$_{\rm{Jupiter}}$  &   Gaussian        &   1.2      &   0.1       \\
Cloud pressure ($\log$(P$_{\rm{cloud}}$))           &  bar                 &   Log-uniform       &  -2        &   1       \\
H$_2$O Mixing Ratio ($\log$(mr$_{\rm{H}_2\rm{O}}$)) &  \nodata             &   Log-uniform      &  -10       &   0      \\
CO Mixing Ratio ($\log$(mr$_{\rm{C}\rm{O}}$))       &  \nodata             &   Log-uniform       &   -10     &   0      \\
CO$_2$ Mixing Ratio ($\log$(mr$_{\rm{C}\rm{O}_2}$)) &  \nodata             &   Log-uniform       &  -10       &   0      \\
CH$_4$ Mixing Ratio ($\log$(mr$_{\rm{C}\rm{H}_4}$)) &  \nodata             &   Log-uniform       &  -10       &   0      \\
Effective temperature (t$_{\rm{int}}$)      &  K                   &   Gaussian   &   1200 &   50  \\
Nearinfrared opacity $\kappa_{\rm{IR}}$      & cgs                   &   Log-uniform   &   -5 &   0  \\
Correlated noise amplitude ($\sigma_{\rm{cor}}$)                 &  W$\cdot$m$^{-2}$    &    Log-uniform     &  -18    &   -15     \\
Correlated noise length ($\lambda_{\rm{cor}}$)                      &  $\mu$m              &   Log-uniform       &   -5     &   0      \\
White noise amplitude ($\sigma_w$)                        &  W$\cdot$m$^{-2}$    &    Log-uniform     &  -18    &   -15     \\
Wavelength shift ($\Delta_\lambda$)         &  $\mu$m              &          Uniform                &   -0.05     &   0.05       \\
Irradiate temperature (t$_{\rm{irr}}$)      &  K                   &   Fixed   &   100 &   \nodata  \\
Distance      &  pc                   &   Fixed   &   10 &   \nodata  
\enddata



\end{deluxetable}

\setlength{\tabcolsep}{-2pt}
\renewcommand{\arraystretch}{1.5}

\begin{deluxetable}{lcccccccccc}
\tabletypesize{\scriptsize}
\tablewidth{0pt}
\tablecaption{Summary of Retrieval Results.\label{tab:mcmc_result}}
\tablehead{
\colhead{\textbf{Parameter}} &
\colhead{\textbf{Unit}} &
\colhead{\textbf{GPI}} &
\colhead{\textbf{CHARIS}} &
\multicolumn{3}{c}{{\bf{OSIRIS}}} &
\multicolumn{4}{c}{{\bf{Combined}}} \\
\colhead{\textbf{}} &
\colhead{\textbf{}} &
\colhead{\textbf{}} &
\colhead{\textbf{}} &
\colhead{\textbf{Abs.}} &
\colhead{\textbf{Norm.}} &
\colhead{\textbf{Norm.}} &
\colhead{\textbf{}} &
\colhead{\textbf{}} &
\colhead{\textbf{}} &
\colhead{\textbf{}} \\
\colhead{\textbf{}} &
\colhead{\textbf{}} &
\colhead{\textbf{}} &
\colhead{\textbf{}} &
\colhead{\textbf{}} &
\colhead{\textbf{Weak}} &
\colhead{\textbf{Strong}} &
\colhead{\textbf{Weak}} &
\colhead{\textbf{Strong}} &
\colhead{\textbf{L+M}} &
\colhead{\textbf{no CO$_2$}} 
}


\startdata
$\log$g                   &  cgs                 &   $3.96^{+0.07}_{-0.06}$        &  $4.00^{+0.06}_{-0.06}$        &  $4.01^{+0.05}_{-0.06}$       &  $3.99^{+0.08}_{-0.08}$      &   3.95  &$3.97^{+0.03}_{-0.03}$&$3.95^{+0.04}_{-0.12}$&$3.99^{+0.06}_{-0.08}$&$3.89^{+0.07}_{-0.06}$     \\
R$_P$                      &  R$_{\rm{Jupiter}}$  &   $1.12^{+0.07}_{-0.06}$        &  $0.95^{+0.03}_{-0.03}$        &  $1.26^{+0.05}_{-0.05}$       &  $1.22^{+0.07}_{-0.08}$      &   0.90  &$1.47^{+0.02}_{-0.02}$&1.20&$1.05^{+0.03}_{-0.03}$ & $0.85^{+0.03}_{-0.02}$   \\
$\log$(P$_{\rm{cloud}}$))          &  bar                 &   $-0.03^{+0.06}_{-0.03}$       &  $-0.59^{+0.15}_{-0.14}$        &  $0.21^{+0.15}_{-0.09}$      &  $-1.26^{+0.38}_{-0.39}$      &   0.05  &$0.26^{+0.02}_{-0.02}$&$0.06^{+0.05}_{-0.05}$&$0.06^{+0.05}_{-0.06}$ & $-0.08^{+0.04}_{-0.06}$   \\
$\log$(mr$_{\rm{H}_2\rm{O}}$) &  \nodata             &   $-2.59^{+0.05}_{-0.04}$       &  $-1.23^{+0.23}_{-0.22}$       &  $-2.95^{+0.07}_{-0.07}$      &  $-0.81^{+0.28}_{-0.33}$     &   $-3.04^{+0.05}_{-0.05}$  &$-2.49^{+0.03}_{-0.03}$&$-2.73^{+0.02}_{-0.03}$&$-2.60^{+0.03}_{-0.03}$ & $-2.62^{+0.04}_{-0.04}$   \\
$\log$(mr$_{\rm{C}\rm{O}}$)       &  \nodata             &   $-2.30^{+0.17}_{-0.21}$       &  $-6.09^{+2.36}_{-2.19}$       &  $-2.79^{+0.25}_{-0.26}$      &  $-0.61^{+0.32}_{-0.40}$     &   $-3.02^{+0.20}_{-0.25}$  &$-2.13^{+0.12}_{-0.13}$&$-2.79^{+0.13}_{-0.16}$&$-2.42^{+0.10}_{-0.10}$ & $-2.54^{+0.08}_{-0.08}$   \\
$\log$(mr$_{\rm{C}\rm{O}_2}$) &  \nodata             &   $-4.39^{+0.16}_{-0.24}$       &  $-6.71^{+2.68}_{-1.89}$       &  $-7.19^{+1.93}_{-1.61}$      &  $-4.16^{+1.89}_{-3.51}$     &   $-3.67^{+0.27}_{-0.24}$   &$-2.94^{+0.06}_{-0.06}$&$-3.06^{+0.06}_{-0.06}$&$-3.12^{+0.08}_{-0.08}$ &  \nodata  \\
$\log$(mr$_{\rm{C}\rm{H}_4}$) &  \nodata             &   $-7.18^{+1.63}_{-1.76}$       &  $-3.24^{+0.19}_{-0.19}$       &  $-7.50^{+1.25}_{-1.39}$      &  $-3.52^{+0.25}_{-0.21}$     &   $-4.92^{+0.12}_{-0.11}$   &$-4.56^{+0.06}_{-0.06}$&$-4.95^{+0.08}_{-0.10}$&$-4.68^{+0.07}_{-0.07}$ &  \nodata   \\
t$_{\rm{int}}$      &  K                   &   $1202^{+34}_{-34}$   &  $1193^{+39}_{-32}$   &  $1220^{+29}_{-34}$  &  $1264^{+33}_{-34}$ &   1213 &$1388^{+7}_{-10}$&1200 &$1390^{+16}_{-22}$ & $1258^{+36}_{-66}$\\
log($\sigma_{\rm{cor}}$)                 &  W$\cdot$m$^{-2}$    &   $-17.08^{+0.68}_{-0.58}$      &  $-16.27^{+0.57}_{-1.14}$      &  $-15.72^{+0.12}_{-0.13}$     &  $-17.24^{+0.58}_{-0.42}$    &   \nodata   &\nodata&\nodata&\nodata&\nodata  \\
log($\lambda_{\rm{cor}}$)                      &  $\mu$m              &   $-2.33^{+1.25}_{-1.73}$       &  $-2.77^{+1.35}_{-1.24}$       &  $-1.44^{+0.17}_{-0.17}$      &  $-1.73^{+1.03}_{-1.22}$     &   \nodata    &\nodata&\nodata&\nodata&\nodata   \\
log($\sigma_w$)                        &  W$\cdot$m$^{-2}$    &   $-17.21^{+0.56}_{-0.48}$      &  $-15.86^{+0.19}_{-0.79}$      &  $-17.29^{+0.57}_{-0.40}$     &  $-17.65^{+0.27}_{-0.19}$    &   \nodata    &\nodata&\nodata&\nodata&\nodata  \\
$\log$($\kappa_{\rm{IR}})$         &  cgs              &   $-2.37^{+0.11}_{-0.10}$                      &  $-1.38^{+0.18}_{-0.17}$       &  $-2.59^{+0.08}_{-0.08}$      &  $-1.97^{+0.66}_{-0.63}$     &   -2.11   &$-3.07^{+0.04}_{-0.04}$&$-2.47^{+0.04}_{-0.05}$&$-2.65^{+0.06}_{-0.05}$&$-2.18^{+0.16}_{-0.12}$    \\
$\Delta_\lambda$         &  nm              &          \nodata                &  \nodata       &  $-0.59^{+0.39}_{-0.40}$      &  $-0.80^{+0.28}_{-0.14}$     &   $-0.54^{+0.04}_{-0.04}$   &$-0.50^{+0.03}_{-0.06}$&$-0.55^{+0.02}_{-0.02}$&$-0.55^{+0.04}_{-0.03}$&$-0.50^{+0.02}_{-0.03}$    \\
t$_{\rm{eff}}$ & K & $1132^{+22}_{-30}$ & $1194^{+16}_{-20}$ & $1127^{+8}_{-16}$ & \nodata & \nodata & $1054^{+7}_{-5}$ & $1102^{+2}_{-2}$ & $1167^{+7}_{-15}$&$1209^{+16}_{-20}$ \\
Luminosity & $10^{-5} \rm{L}_\odot$ &  $1.94^{+0.10}_{-0.06}$ & $1.76^{+0.05}_{-0.05}$ & $2.40^{+0.11}_{-0.20}$ & \nodata & \nodata & $2.55^{+0.04}_{-0.04}$ & $2.02^{+0.01}_{-0.01}$ & $1.96^{+0.04}_{-0.03}$& $1.45^{+0.04}_{-0.03}$
\enddata


\end{deluxetable}

\end{CJK*}
 
\end{document}